\documentclass[11pt,a4paper]{article}

\usepackage{amsmath,amssymb}
\usepackage{epsfig,graphicx}
\usepackage{subfigure}
\usepackage{graphicx}
\usepackage{rotating}
\usepackage{cancel}
\usepackage{bm}
\usepackage{color}
\usepackage[svgnames]{xcolor}
\usepackage{comment}
\usepackage{cite}
\usepackage{psfrag}
\usepackage{slashed}
\usepackage{soul}
\usepackage{tikz}
\usepackage{bbold}
\usepackage{slashed}
\usepackage[normalem]{ulem}

\numberwithin{equation}{section}

\renewcommand\[{\left[}

\newcommand{\exclude}[1]{}

\def\beq{\begin{equation}}
\def\eeq{\end{equation}}

\usepackage[colorlinks=true,linkcolor=black,urlcolor=DarkBlue,citecolor=teal]{hyperref}

\usepackage[english]{babel}
\addto\extrasenglish{%
  \def\sectionautorefname{Section}%
  \def\subsectionautorefname{Section}%
  \def\equationautorefname~#1\null{Equation~(#1)\null}
}

\usepackage{etoolbox}
\makeatletter
\patchcmd{\hyper@makecurrent}{%
    \ifx\Hy@param\Hy@chapterstring
        \let\Hy@param\Hy@chapapp
    \fi
}{%
    \iftoggle{inappendix}{
        \@checkappendixparam{chapter}%
        \@checkappendixparam{section}%
        \@checkappendixparam{subsection}%
        \@checkappendixparam{subsubsection}%
        \@checkappendixparam{paragraph}%
        \@checkappendixparam{subparagraph}%
    }{}%
}{}{\errmessage{failed to patch}}

\newcommand*{\@checkappendixparam}[1]{%
    \def\@checkappendixparamtmp{#1}%
    \ifx\Hy@param\@checkappendixparamtmp
        \let\Hy@param\Hy@appendixstring
    \fi
}
\makeatother

\usepackage{xstring}
\newcommand{\subautoref}[1]{%
  \IfBeginWith{#1}{app:}%
    {\hyperref[#1]{Appendix~\ref*{#1}}}%
    {\hyperref[#1]{Section~\ref*{#1}}}%
}

\newtoggle{inappendix}
\togglefalse{inappendix}

\apptocmd{\appendix}{\toggletrue{inappendix}}{}{\errmessage{failed to patch}}

\begin{document}
\title{\huge{\textbf{Testing Antimatter Couplings with Spectroscopy}}
\\[0.3cm]}

\author{Joerg Jaeckel and Lucas Puetter\\
\small{\em Institut f\"ur Theoretische Physik, Universit\"at Heidelberg,}\\
\small{ \em Philosophenweg 16, 69120 Heidelberg, Germany} }

\date{}
\maketitle

\begin{abstract}
\noindent We investigate how scalar-mediated potentials with Lorentz-violating couplings to Standard Model fermions affect spectroscopic observables in atoms and highly charged ions. Suitable combinations of an ordinary scalar and a time-like component of a Lorentz violating vector coupling allow for a split into ``matter'' and ``antimatter'' couplings, at least in the non-relativistic limit. By considering hydrogen and antihydrogen spectra, we access both matter and antimatter couplings. While relativistic effects alone lift degeneracies in ordinary hydrogen, providing indirect access to antimatter couplings, comparisons with antihydrogen measurements lead to significantly improved sensitivity to the antimatter couplings. In highly charged ions, enhanced relativistic effects further amplify the sensitivity, compensating for reduced experimental precision and larger theoretical uncertainties. We obtain the strongest bounds to date for scalar masses $m_\phi \gtrsim 400\:\mathrm{keV}$. For comparison, we estimate astrophysical constraints on the same parameter space, providing strong bounds even on antimatter couplings, despite stars being predominantly composed of matter.
\end{abstract}

\newpage

\tableofcontents
\newpage


\section{Introduction}\label{sec:intro}

Antimatter is still surrounded by a light whiff of the mysterious. Many years after its prediction~\cite{Dirac:1931kp} and discovery~\cite{Anderson:1933mb}, there is a multitude of experiments~\cite{AEGISProto:2008qxw,ALPHA:2010wjo,ASACUSA:1997vux,Ulmer:2011pra,GBAR:2011fij,PUMA:2022ngr} aiming to pierce the veil by probing it at the highest levels of precision. This includes measurements of masses of antimatter particles~\cite{BASE:2022yvh} as well as precision determinations of $g$-factors~\cite{VanDyck:1987ay,BASE:2016yuo}. One of the most precise tests, and one that we will use in the present paper, is spectroscopy of antihydrogen~\cite{ALPHA:2020rbx,Baker:2025ehs}.

At the same time, feebly interacting particles (FIPs) of a ``dark sector'' beyond the Standard Model are both theoretically well motivated and are coming under increasing experimental scrutiny (see, e.g.~\cite{Jaeckel:2010ni,Safronova:2017xyt,Beacham:2019nyx,Agrawal:2021dbo,Antypas:2022asj,Adams:2022pbo,Antel:2023hkf,Cong:2024qly,Albertus:2026fbe,Arza:2026rsl} for some reviews). Importantly, feeble interactions with Standard Model particles make it much harder to test these new particles experimentally and even light ones may have been overlooked. Yet, their low mass and their feebly interactions may have a simultaneous origin in some underlying mechanism, operating at very high energy scales. For example, both the coupling and the mass of axion-like particles are usually suppressed by the same scale which gives rise to their very existence as pseudo-Goldstone bosons of an underlying shift symmetry. Small masses, in turn, enable a wide variety of new and fast developing experimental methods to come into play (cf., again~\cite{Jaeckel:2010ni,Safronova:2017xyt,Beacham:2019nyx,Agrawal:2021dbo,Antypas:2022asj,Adams:2022pbo,Antel:2023hkf,Cong:2024qly,Albertus:2026fbe,Arza:2026rsl}), making light FIPs a very promising tool to probe the dark sector.

Now, what if a difference between matter and antimatter is only carried by a relatively light, feebly interacting particle? To describe this situation, we need a FIP that interacts via CPT-violating interactions with the Standard Model particles. By the CPT theorem~\cite{Luders:1954zz,Pauli1988},\footnote{See~\cite{Blum:2022eol} for a detailed history including also J.S. Bell's involvement.} the latter also implies Lorentz violating interactions (see~\cite{Kostelecky:1988zi,Kostelecky:1991ak,Gambini:1998it,Mocioiu:2000ip,Carroll:2001ws,Alfaro:2001rb,Kostelecky:2002ca,Altschul:2005mu,Ferrero:2009jb} for some motivations of Lorentz symmetry breaking). A suitable model is the one given in~\cite{Altschul:2012xu} (building on~\cite{Altschul:2006jj,Ferrero:2011yu}), further analyzed in~\cite{Altschul:2014gqa} and more recently in~\cite{Carenza:2025jwn}. This includes a scalar boson interacting with SM fermions via all possible Lorentz structures up to dimension 4. The relevant structures will be given below in \autoref{sec:model}. This combines Lorentz violation~\footnote{For spectroscopic tests of the SME see, e.g.~\cite{Kostelecky:2015nma}.} with the inclusion of a new light degree of freedom and thereby goes beyond the Standard Model Extension (SME)~\cite{Colladay:1996iz, Colladay:1998fq,Kostelecky:2003fs} that is usually employed to study Lorentz violation but does not include any new degrees of freedom. 

The new degrees of freedom lead to new non-pointlike interaction potentials~\cite{Altschul:2012xu}, or they can be produced as real final state particles, notably in suitable astrophysical environments~\cite{Carenza:2025jwn,Lucenteprep} (see~\cite{Raffelt:1996wa} for a general introduction to astrophysical tests of FIPs). In the present paper, we will focus on testing the former via precision measurements on simple atomic and ionic systems and make rough estimates of the latter, leaving a more advanced calculation to future work~\cite{Lucenteprep}.

Focusing on the distinction of matter vs antimatter, it makes sense to elucidate this more clearly by identifying structures that couple our scalar of interest only to matter or antimatter, respectively. One can then, e.g. test ALP dark matter coupled purely to antimatter~\cite{Smorra:2019qfx} or new ALP or (axial-)vector forces between antimatter particles~\cite{Cong:2025wge}. Steps in this direction were taken already in~\cite{Altschul:2012xu}, elucidating the different potentials for matter and antimatter, depending on the coupling structure, and in~\cite{Haagner,Souissi,Souissi2} by looking at potentially suitable Lorentz structures.
In \autoref{sec:model}, we give the relevant combinations for a simple scalar matter/antimatter interaction. Based on a Lagrangian in terms of Lorentz structures, it becomes clear that such a pure matter/antimatter coupling is actually an artefact of the extreme non-relativistic limit. Taking relativistic corrections into account, a purely antimatter coupled scalar also interacts with matter, albeit with velocity suppressed strength.\footnote{Here, we only consider the simplest Lorentz violating interaction up to operator dimension 4. Inclusion of higher dimensional operators may allow for additional suppression, but generally at the price of more problematic UV behaviour.} This is one of the main messages of our paper: {\emph{We can test antimatter couplings with matter experiments.}}

The paper follows the following rough path. In the next \autoref{sec:theory}, we specify the model, discuss combinations that provide ``pure'' matter/antimatter couplings, but also the velocity dependent interactions with the respective antimatter/matter parts. In this context, we also give the interaction potentials relevant in atomic systems, and mention the frame dependence inherent in Lorentz violating theories. \autoref{sec:analysis_framework} then describes our analysis framework. This is applied in \autoref{sec:results} to precision spectroscopy of hydrogen, antihydrogen as well as hydrogen-like (1 electron) highly charged ions, giving sensitive tests of both matter and antimatter couplings. \autoref{sec:other} then complements this with other bounds, notably an estimate of ones arising from astrophysical systems. Brief conclusions are given in \autoref{sec:conclusions}.


\section{Theoretical setup}\label{sec:theory}


\subsection{Model}\label{sec:model}
While the effective field theory approach of the SME~\cite{Colladay:1996iz, Colladay:1998fq,Kostelecky:2003fs} does not include new particles at accessible energies, we consider a model in which a new, light scalar is coupled to SM fermions. The corresponding Yukawa couplings are assumed to be the only source of Lorentz violation (LV), giving rise to Lorentz-violating potentials at low energies \cite{Altschul:2006jj,Ferrero:2011yu,Altschul:2012xu}.

In addition to the SM, we consider a Lagrangian \cite{Altschul:2012xu}
\begin{align}
    \mathcal L_Y
    &=
    \frac 1 2\left(\partial_\mu\phi\right)\left(\partial^\mu\phi\right)-\frac 1 2 m_\phi^2\phi^2-\phi\sum_F\bar\psi_FG_F\psi_F,\label{eq:Lagrangian}\\
    G_F&=g_F+ig'_F\gamma_5+I_\mu^F\gamma^\mu+J_\mu^F\gamma_5\gamma^\mu+\frac 1 2L_{\mu\nu}^F\sigma^{\mu\nu},
    \label{eq:GF_coupling}
\end{align}
where $m_\phi$ is the mass of the scalar field $\phi$. $F$ indicates the SM fermion species that the new scalar couples to. The coefficients $g_F$ and $g_F'$ correspond to the conventional scalar and pseudoscalar couplings, respectively, while the vector, pseudovector and second-rank tensorial coefficients are given by $I_\mu^F$, $J_\mu^F$ and $L_{\mu\nu}^F$. All coefficients are dimensionless and the latter three lead to LV. Importantly, for our purposes, the vector ($I_\mu^F$) and pseudovector ($J_\mu^F$) terms are also CPT-violating, as required to make a distinction between matter and antimatter.

While, in principle, all terms in \autoref{eq:GF_coupling} contribute to the coupling $G_F$, we restrict our discussion to the scalar and vector couplings for simplicity and in particular choose $I_j^F=0$ in a specific frame, i.e. $G_F=g_F+I_0^F\gamma^0$ in that frame. Aside from simplicity, our choice is motivated by the fact that this is the simplest set of couplings that, suitably combined, can make the scalar coupled purely to matter or antimatter in the non-relativistic limit. As we will see below, in \autoref{sec:potential}, and roughly speaking, in this limit, the coupling is given by
\begin{equation}
(g_F\pm I_{0}^F)
\end{equation}
to matter (+) and antimatter (-) components, respectively. Therefore, choosing $g=-I_{0}$ is a ``pure'' antimatter coupling of the scalar.

We leave the discussion of the other coefficients to future work.


\subsection{Frame Dependence}
\label{sec:frame}
Dealing with a Lorentz violating theory, we have to make a choice of frame. One natural choice is to select the rest frame of the cosmic microwave background (CMB) as the reference frame for our couplings. In this frame, we assume $g_F^\mathrm{CMB}\equiv g_F\neq 0$, $I_{0,\mathrm{CMB}}^F\equiv I_0^F\neq 0$ and $I_{j,\mathrm{CMB}}^F\equiv I_j^F=0$. The Lorentz-violating framework described in the previous section is invariant under observer Lorentz transformations. Due to the relative velocity of the solar system with respect to the CMB frame, which has a magnitude of $v_\mathrm{CMB}\sim 10^{-3}$ \cite{Planck:2013kqc}, the laboratories in which, e.g., spectroscopic experiments are performed observe transformed values of the couplings. While $g_F$ is a Lorentz-scalar and hence remains the same in all frames, the vector components transform as
\begin{align}
    I_{0,\mathrm{lab}}^F
    =
    \gamma_\mathrm{CMB}I_0^F,\qquad
    I_{j,\mathrm{lab}}^F=-\gamma_\mathrm{CMB}v_{\mathrm{CMB},j}I_0^F,
\end{align}
where $v_{\mathrm{CMB},j}$ describe the components of the velocity of the solar system in the CMB frame and $\gamma_\mathrm{CMB}=(1-v_\mathrm{CMB}^2)^{-1/2}$. As $v_\mathrm{CMB}$ dominates over the velocity of the Earth around the sun by an order of magnitude, we disregard the latter in the forthcoming analysis and note that this would lead to a small sidereal variation of the corresponding observables.


\subsection{Potential}
\label{sec:potential}
The interaction potential between fermions is amended by the Yukawa interaction in \autoref{eq:Lagrangian}. We are interested in atomic and ionic systems that contain nucleons and electrons. For the former -- here vertex ($a$) -- we directly apply the non-relativistic approximation (to the lowest order of interest). For the latter -- vertex ($b$) -- we use relativistic Dirac wave functions. The resulting particle-particle potential in a given frame is then given by (see~\cite{Altschul:2012xu} and \autoref{sec:app_potential} for a detailed derivation of all considered potentials)
\begin{align}
\label{eq:potential_hydrogen}
    V_\mathrm{frame}^{ab}(r)=\left(g_a+I_{0,\mathrm{frame}}^a\right)\left[g_b\gamma^0+I_{0,\mathrm{frame}}^b\right]f(r).
\end{align}
Here, it is understood that the term in square brackets is to be contracted with the corresponding spinors of particle $b$ to obtain the interaction energy. The radial function is given by the Yukawa potential function,
\begin{align}
    f(r)=-\frac{e^{-m_\phi r}}{4\pi r}.
\end{align}
The corresponding antiparticle-antiparticle potential is
\begin{align}
\label{eq:potential_antihydrogen}
    V_\mathrm{frame}^{\bar a\bar b}(r)
    =
    -\left(g_a-I_{0,\mathrm{frame}}^a\right)\left[g_b\gamma^0+I_{0,\mathrm{frame}}^b\right]f(r),
\end{align}
where the term in square brackets needs to be contracted with antiparticle spinors of $b$. Note the crucial sign change between the Lorentz-symmetric and Lorentz-violating couplings of particle $a$.
Hence, as already pointed out in~\cite{Altschul:2012xu}, these potentials provide a distinct coupling for particles and antiparticles in the non-relativistic limit, given by $(g_a+I_0^a)$ and $(g_a-I_0^a)$, respectively.  
The same relative signs appear for particle $b$ when contracting with the appropriate spinors, including important relativistic corrections originating from the small spinor components. 

In a general frame, there are additional potential terms, in particular including the spatial components $I_{j,\mathrm{frame}}^F$. The full potential valid in any frame is given in \autoref{sec:app_full_potential}. However, the additional terms are subdominant in our analysis since they are either strongly velocity-suppressed, vanish in first-order perturbation theory due to parity, such that the leading term is quadratic in the couplings, or couple to spin only, leading to hyperfine-structure effects which are not considered in the forthcoming analysis.

Assuming the couplings in the CMB frame, as described in \autoref{sec:frame}, and a stationary laboratory on Earth in which the probed atoms are approximately motionless, one finds the particle potential in terms of the fundamental CMB couplings to be
\begin{align}
    V_\mathrm{lab}^{ab}(r)
    &=
    \left(g_a+\gamma_\mathrm{CMB}I_0^a\right)\left[g_b\gamma^0+\gamma_\mathrm{CMB}I_0^b\right]f(r)\nonumber\\
    &=
    \left(c_+g_a^+-c_-g_a^-\right)\left[c_+\mathcal G_b^+-c_-\mathcal G_b^-\right]f(r)\,.
    \label{eq:potential_hydrogen_CMB}
\end{align}
Here, we defined $c_\pm\equiv (\gamma_\mathrm{CMB}\pm1)/2$ as well as the matter and antimatter couplings,
\begin{equation}
    g_a^\pm\equiv g_a\pm I_0^a
\end{equation}
for the non-relativistic and 
\begin{equation}
    \mathcal G_b\equiv g_b\gamma^0\pm I_0^b
\end{equation}    
for the relativistic particle. Expanding $\gamma_\mathrm{CMB}$ to leading order in $v_\mathrm{CMB}$, we find 
\begin{align}
    c_+\simeq 1+v_\mathrm{CMB}^2/4,\qquad c_-\simeq v_\mathrm{CMB}^2/4,
\end{align}
indicating a $v_\mathrm{CMB}^2$-suppression of the non-relativistic antiparticle couplings $g_F^-$ with respect to the particle couplings $g_F^+$. The antiparticle potential in the laboratory frame can be derived analogously. It is obtained from the particle potential by exchanging the sign of $I_0^a$ and multiplying with a global sign. If an experiment is performed on atoms moving with a velocity $v_\mathrm{exp}$ with respect to the lab frame, we have to replace $v_\mathrm{CMB}$ with the total velocity with respect to the CMB.
In many cases, there is a clear hierarchy in the velocities, such that we can approximate the combined velocity simply by using the largest one.


\subsection{Dirac Equation for (Anti)hydrogen and Hydrogenic Ions}
\label{sec:Dirac}

To be explicit and to set our conventions, let us briefly recall the solutions of the Dirac equation for the electron in hydrogen and hydrogenic ions, see, e.g.~\cite{rose1961relativistic}. The relativistic Dirac equation for the hydrogen atom is given by
\begin{align}
\label{eq:Dirac}
    \left(-i\boldsymbol{\alpha}\cdot\nabla+\beta m_e+V(r)\right)\psi(\boldsymbol{r})=E\psi(\boldsymbol{r})
\end{align}
with $\boldsymbol\alpha$ and $\beta$ the usual Dirac matrices and $V(r)=-Z\alpha/r$ the Coulomb potential. In spherical coordinates, the solution can be written in the form~\cite{rose1961relativistic}
\begin{align}
\label{eq:spinor_hydrogen}
    \psi_\mathrm{H}(\boldsymbol r)
    =
    \begin{pmatrix}
        G(r)\mathcal Y_\kappa^m(\theta,\phi)\\
        iF(r)\mathcal Y_{-\kappa}^m(\theta,\phi)
    \end{pmatrix},
\end{align}
where $G(r)$ and $F(r)$ are the real-valued radial functions corresponding to the large and small components of a positive-energy solution. $\mathcal Y_\kappa^m(\theta,\phi)$ are the angular two-component spinors, which satisfy
\begin{align}
\label{eq:angular_orthonormality}
    \int\mathrm d\Omega\left(\mathcal Y_{\kappa_a}^{m_a}\right)^\dagger\mathcal{Y}_{\kappa_b}^{m_b}
    =
    \delta_{j_aj_b}\delta_{m_am_b}\delta_{\ell_a\ell_b},
\end{align}
where $\kappa\equiv (-1)^{j+\ell+1/2}\left(j+\frac 1 2\right)$ is the angular quantum number, with $j$, $m$ and $\ell$ the total, magnetic, and orbital angular momentum quantum numbers. The charge-conjugate $\psi^c=i\gamma^2\psi^*$ is also a solution of the same Dirac equation in \autoref{eq:Dirac}. With the above choice of $\psi_\mathrm{H}$, we have identified the positive-energy solution with hydrogen, such that the solution for antihydrogen is given by the negative energy-solution and is hence found by charge conjugation of the hydrogen solution up to unphysical phases,
\begin{align}
\label{eq:spinor_antihydrogen}
    \psi_{\bar{\mathrm{H}}}(\boldsymbol r)
    =
    \begin{pmatrix}
        -iF(r)\mathcal Y_{-\kappa}^{-m}(\theta,\phi)\\
        G(r)\mathcal Y_{\kappa}^{-m}(\theta,\phi)
    \end{pmatrix}\,.
\end{align}
Here, the radial functions are the same as for the hydrogen atom and again $G(r)$ is the large and $F(r)$ the small component. See \autoref{sec:app_Dirac} for a more detailed derivation.

We use the expressions from \cite{drake2023springer} for $G(r)$ and $F(r)$, corresponding to positive-energy bound states of hydrogenic ions with nuclear charge $Z$. The energy eigenvalues are given by
\begin{align}
    E_{n_r,\kappa}
    =
    m_e\left[1-\frac{(Z\alpha)^2}{(n_r+1+\gamma)^2}\right]^{-1/2}
\end{align}
with $m_e$ the electron mass, $n_r=n-\vert\kappa\vert$, where $n$ is the principal quantum number, and $\gamma=-1+\left(\kappa-(Z\alpha)^2\right)^{1/2}$. 
Note that, in the conventions of \cite{drake2023springer}, one could directly use the hydrogen spinor from \autoref{eq:spinor_hydrogen} for antihydrogen by assuming negative energy as well as opposite signs for $\kappa$ and $Z$ in the definitions of $G(r)$ and $F(r)$. Our approach is complementary and intuitive by making the connection with the charge conjugate.

Expanding the energy to first order in powers of $Z\alpha$, we can approximate
\begin{align}
    G
    &\propto
    (m_e+E_{n_r,\kappa})^{1/2}
    \approx
    m_e^{1/2}\left(1+1-\frac 1 2\frac{(Z\alpha)^2}{n_r+1+\gamma}\right)^{1/2}
    \sim
    \sqrt{2m_e},\\
    F
    &\propto
    (m_e-E_{n_r,\kappa})^{1/2}
    \approx
    m_e^{1/2}\left(1-1+\frac 1 2\frac{(Z\alpha)^2}{n_r+1+\gamma}\right)^{1/2}
    \sim 
    \frac{\sqrt{2m_e}}{2}Z\alpha.
\end{align}
Thus, $F$ is suppressed in comparison to $G$ by $Z\alpha/2$, $F/G\sim Z\alpha/2$. Applying the virial theorem, one can show that $Z\alpha\sim v$, leading to a $v/2$-suppression of the small component. We remark that the suppression of the small spinor components by $v$ in comparison to the large ones is a general feature also valid for free particle spinors, which we will also use in the discussion of astrophysical bounds in \autoref{sec:astro_bounds}.


\subsection{Effects of the New Potential on Energy Levels}
\label{sec:energy_correction}
We calculate the energy correction to the states of hydrogen\footnote{Note that all hyperfine states of a given energy level receive the same energy shift.} due the the presence of the new potential in first-order perturbation theory. For hydrogen, we use the potential from \autoref{eq:potential_hydrogen_CMB} with $a=p$ and $b=e^-$ the proton and electron, respectively. For a quantum state $Q$ with quantum numbers $n,\ell,j,m$, the energy correction is given by
\begin{align}
    \Delta E_\mathrm{H}^Q
    &=
    \left(g_p+\gamma_\mathrm{CMB}I_0^p\right)\left\langle n\ell jm\left\vert\left[g_e\gamma^0+\gamma_\mathrm{CMB}I_0^e\right]f(r)\right\vert n\ell jm\right\rangle\nonumber\\
    &=
    -\frac{c_+g_p^+-c_-g_p^-}{4\pi}\left[g_e^+\left(c_+I_G+c_-I_F\right)-g_e^-\left(c_-I_G+c_+I_F\right)\right]\label{eq:Delta_E_H_c}\\
    &\simeq
    -\frac{g_p^+-\frac{v_\mathrm{CMB}^2}{4}g_p^-}{4\pi}\left[g_e^+\left(I_G+\frac{v_\mathrm{CMB}^2}{4}I_F\right)-g_e^-\left(\frac{v_\mathrm{CMB}^2}{4}I_G+I_F\right)\right]\label{eq:Delta_E_H_v},
\end{align}
where we used the hydrogen spinor in spherical coordinates from \autoref{eq:spinor_hydrogen} and the orthonormality relation for the angular part from \autoref{eq:angular_orthonormality}. We also approximated $c_+\approx 1$ and $c_-\approx\frac{v_\mathrm{CMB}^2}{4}$ in the last step and defined
\begin{align}
    I_G
    \equiv
    \int\mathrm dr\:re^{-m_\phi r}G(r)^2,\qquad
    I_F
    \equiv
    \int\mathrm dr\:re^{-m_\phi r}F(r)^2.
\end{align}

As usual in hydrogen-like systems, the correction only contains products of electron and proton couplings. Therefore, only these products can be constrained.\footnote{See~\cite{Cong:2026kuv,Abdullin:2026zdi,Moretti} for analyses of systems with more than one electron which also give access to the electron coupling by itself.}

As already alluded to, $F$ is velocity-suppressed compared to $G$ and hence 
\begin{equation}
I_{F}\sim \frac{v^2}{4} I_{G},
\end{equation}
where $v$ is the velocity of the electron.
Accordingly, the largest term in the energy correction is proportional to $g_p^+g_e^+$, i.e. the product of the particle couplings. However, at a velocity-suppressed level also the antiparticle couplings appear in the energy correction.

Let us briefly consider the relevant hierarchies. In the centre of mass frame of the hydrogen atom, the proton moves with a velocity much smaller than $v_\mathrm{CMB}$. Therefore, the $v_\mathrm{CMB}^2$-suppressed term dominates in accessing the antiproton coupling. On the other hand, the typical electron velocity is larger than the velocity relative to the CMB, $v\sim\alpha>v_\mathrm{CMB}$. Thus, there are two competing contributions to the term including the positron coupling $g_e^-$. Due to the velocity hierarchy, one finds $I_F>\frac{v_\mathrm{CMB}^2}{4}I_G$. This discussion shows that the non-relativistic positron and antiproton couplings, despite being suppressed by $v^2\sim\alpha^2$ or $v_\mathrm{CMB}^2$, are accessible even in hydrogen spectroscopy experiments, albeit with velocity suppressed sensitivity. Note that this hierarchy is valid only for approximately motionless atoms in the laboratory frame. If $v_\mathrm{exp}\gtrsim v_\mathrm{CMB}$, the corresponding velocities need to be replaced, such that one may find $I_F<\frac{v_\mathrm{exp}^2}{4}I_G$.

In the case of antihydrogen, we use the corresponding antiparticle potential from \autoref{eq:potential_hydrogen_CMB} with $\bar a=\bar p$ the antiproton and $\bar b =e^+$ the positron. Using the antihydrogen spinor from \autoref{eq:spinor_antihydrogen}, we find the energy correction
\begin{align}
    \Delta E_{\bar{\mathrm{H}}}^Q
    &=
    -\left(g_p-\gamma_\mathrm{CMB}I_0^p\right)\left\langle n\ell jm\left\vert\left[g_e\gamma^0+\gamma_\mathrm{CMB}I_0^e\right]f(r)\right\vert n\ell jm\right\rangle\nonumber\\
    &=
    -\frac{c_+g_p^--c_-g_p^+}{4\pi}\left[g_e^-\left(c_+I_G+c_-I_F\right)-g_e^+\left(c_-I_G+c_+I_F\right)\right]\label{eq:Delta_E_Hbar_c}\\
    &\simeq
    -\frac{g_p^--\frac{v_\mathrm{CMB}^2}{4}g_p^+}{4\pi}\left[g_e^-\left(I_G+\frac{v_\mathrm{CMB}^2}{4}I_F\right)-g_e^+\left(\frac{v_\mathrm{CMB}^2}{4}I_G+I_F\right)\right],
\end{align}
where, again, the angular dependence is trivial. In this case, the dominant energy correction is proportional to the antimatter couplings $g_p^-g_e^-$. As in the case of hydrogen, the charge-conjugate couplings are accessible but again suppressed by $v^2$ or $v_\mathrm{CMB}^2$ compared to the antiparticle couplings.


\section{Analysis Framework}
\label{sec:analysis_framework}
We constrain the allowed parameter space for the scalar and Lorentz-violating time-like component vector coupling by comparing experimental spectroscopic measurements with corresponding theory values and performing a least-squares fit. After presenting and discussing the experimental and theoretical values, we briefly describe the analysis procedure.

\begin{table}[t]
    \centering
    \resizebox{\columnwidth}{!}{%
    \begin{tabular}{c c c c c c c}
    \hline\hline
        Index & Transition & Velocity & Theory (MHz) & Experiment (MHz) & Refs.\\
        \hline
         1&$1\mathrm S_{1/2}-2\mathrm S_{1/2}$ (H)&$\sim 0$&$2\:466\:061\: 413.182\:1\:(44)$& $2\:466\:061\: 413.187\:035\:(10)$&\cite{Parthey:2011lfa} \\
         2&$1\mathrm S_{1/2}-3\mathrm S_{1/2}$ (H)&$\sim 0$&$2\:922\:743\:278.658\:5\:(51)$& $2\:922\:743\:278.665\:79\:(72)$&\cite{Grinin:2020txk} \\
         3&$2\mathrm P_{1/2}-2\mathrm S_{1/2}$ (H)&$0.01$&$1\:057.832\:30\:(24)$& $1\:057.829\:8\:(32)$&\cite{Bezginov:2019mdi}\\
         4&$2\mathrm S_{1/2}-2\mathrm P_{3/2}$ (H)&$0.014$&$9\:911.209\:21\:(24)$& $9\:911.200\:(12)$&\cite{Hagley1994}\\
         5&$2\mathrm S_{1/2}-4\mathrm S_{1/2}$ (H)&$\sim 0$&$616\:520\:150.616\:8\:(10)$& $616\:520\:150.635\:(10)$&\cite{Weitz:1995zz,Parthey:2011lfa}\\
         6&$2\mathrm S_{1/2}-4\mathrm P_{1/2}$ (H)&$\sim 0$&$616\:520\:017.540\:9\:(10)$& $616\:520\:017.538\:7\:(30)$&\cite{Beyer:2017gug}\\
         7&$2\mathrm S_{1/2}-4\mathrm P_{3/2}$ (H)&$\sim 0$&$616\:521\:388.670\:7\:(10)$& $616\:521\:388.670\:9\:(30)$&\cite{Beyer:2017gug}\\
         8&$2\mathrm S_{1/2}-6\mathrm P_{1/2}$ (H)&$\sim 0$&$730\:689\:977.769\:6\:(12)$& $730\:689\:977.770\:88\:(69)$&\cite{Maisenbacher:2026nau}\\
         9&$2\mathrm S_{1/2}-6\mathrm P_{3/2}$ (H)&$\sim 0$&$730\:690\:384.029\:3\:(12)$& $730\:690\:384.030\:74\:(66)$&\cite{Maisenbacher:2026nau}\\
         10&$2\mathrm S_{1/2}-8\mathrm D_{5/2}$ (H)&$\sim 0$&$770\:649\:561.562\:5\:(13)$& $770\:649\:561.570\:9\:(20)$&\cite{Brandt:2021yor}\\[2mm]
         11&$1\mathrm S_{1/2}-2\mathrm S_{1/2}$ ($\bar{\mathrm{H}}$,$*_1$)&$\sim 0$&$2\:466\:061\: 413.182\:1\:(44)$& $2\:466\:061\:413.189\:2\:(59)$&\cite{Baker:2025ehs,Rasmussen:2017pyn}\\
         12&$1\mathrm S_{1/2}-2\mathrm S_{1/2}$ ($\bar{\mathrm{H}}$,$*_2$)&$\sim 0$&$2\:466\:061\: 413.182\:1\:(44)$& $2\:466\:061\:413.186\:2\:(63)$&\cite{Baker:2025ehs,Rasmussen:2017pyn}\\
         13&$1\mathrm S_{1/2}-2\mathrm S_{1/2}$ ($\bar{\mathrm{H}}$,$*_3$)&$\sim 0$&$2\:466\:061\: 413.182\:1\:(44)$& $2\:466\:061\:413.180\:0\:(92)$&\cite{Baker:2025ehs,Rasmussen:2017pyn}\\
         14&$2\mathrm P_{1/2}-2\mathrm S_{1/2}$ ($\bar{\mathrm{H}}$)&$\sim 0$&$1\:057.832\:30\:(24)$&$1\:046\:(35)$&\cite{ALPHA:2020rbx}\\
         15&$2\mathrm P_{1/2}-2\mathrm P_{3/2}$ ($\bar{\mathrm{H}}$)&$\sim 0$&$10\:969.041\:514\:(6)$&$10.88\:(19)\times 10^3$ &\cite{ALPHA:2020rbx}\\
         \hline\hline
    \end{tabular}
    }
    \caption{Hydrogen (H) and antihydrogen ($\bar{\mathrm{H}}$) transitions considered in the analysis with respective theoretical and experimental values and corresponding references for the experimental measurements. The theoretical values follow from the theoretical framework of \cite{Mohr:2024kco}. All values correspond to the hyperfine centroids. The antihydrogen measurements denoted with $*_i$ correspond to three different measurements, which differ by run of the experiment or the measured hyperfine transition \cite{Baker:2025ehs}. The velocities are taken in the laboratory frame. Atomic velocities of experiments performed at cryogenic temperatures are denoted by $\sim 0$ as they are irrelevant in comparison to $v_\mathrm{CMB}$. See text for details.}
    \label{tab:data}
\end{table}

\subsection{Experimental Input}
The high-precision spectroscopic results from hydrogen and antihydrogen experiments used in the analysis are listed in \autoref{tab:data}, together with their corresponding references. We use transition frequencies between the hyperfine-structure centroids, i.e. the degeneracy-weighted average energy of all hyperfine sublevels belonging to the corresponding state, since the energy corrections discussed in \autoref{sec:energy_correction} arise at the fine-structure level. This also facilitates a direct comparison with theoretical values, without requiring the inclusion of nuclear-structure effects in the hyperfine calculations. 

Some of the references given in \autoref{tab:data} do not provide the values for the fine-structure centroid directly and must therefore be converted. Starting with the hydrogen transitions, the $2\mathrm{S}_{1/2}-4\mathrm S_{1/2}$ transition frequency is obtained by combining the quantity $(E_{4\mathrm S_{1/2}}-E_{2\mathrm S_{1/2}})-\frac 1 4(E_{2\mathrm S_{1/2}}-E_{1\mathrm S_{1/2}})$ from \cite{Weitz:1995zz} with the measured $1\mathrm{S}_{1/2}-2\mathrm S_{1/2}$ transition frequency from \cite{Parthey:2011lfa}. In references \cite{Beyer:2017gug,Maisenbacher:2026nau}, the transition frequencies are given in terms of the hyperfine states for the $2\mathrm{S}_{1/2}-4\mathrm P_{j}$ and $2\mathrm{S}_{1/2}-6\mathrm P_{j}$ transitions, respectively. We use the formulae provided in the respective supplemental material to determine the hyperfine centroids. In all cases, uncertainties are propagated and the resulting correlations are taken into account. As we will see, the measurements of the $2\mathrm P_{1/2}-2\mathrm S_{1/2}$ \cite{Bezginov:2019mdi} and $2\mathrm S_{1/2}-2\mathrm P_{3/2}$ \cite{Hagley1994} transitions, performed with a fast beam of hydrogen atoms at velocities close to 1\% of the speed of light, help to lift degeneracies, improving the bounds set in combination with antihydrogen spectroscopy by several orders of magnitude and allowing for bounds to be placed by hydrogen only.

The latest antihydrogen measurements of the $1\mathrm{S}_{1/2}-2\mathrm S_{1/2}$ transition by the ALPHA experiment \cite{Baker:2025ehs} correspond to transitions between hyperfine states in a magnetic field. The magnetic-field correction is computed following \cite{Rasmussen:2017pyn}, yielding transition frequencies between hyperfine states at zero magnetic field. Although the ALPHA experiment has measured the hyperfine splittings of the $1\mathrm{S}_{1/2}$ \cite{ALPHA:2026lbu} and $2\mathrm S_{1/2}$ \cite{Baker:2025ehs} states in antihydrogen, we instead use the respective hydrogen values \cite{Karshenboim:2005iy,Bullis:2023big}. This avoids introducing a dominant uncertainty from the hyperfine splitting that would exceed the transition uncertainty of interest. The resulting transition frequencies between the hyperfine-centroids are the ones listed in \autoref{tab:data}. As before, uncertainties are propagated and relevant correlations are included. The Lamb shift and $2\mathrm P$ fine-structure splitting for antihydrogen are given in \cite{ALPHA:2020rbx} for zero magnetic field between the hyperfine centroids and can therefore be used without modification.

\subsection{Theory Input}
\label{sec:theo_values}
To investigate possible contributions from new physics, we compare the experimental values with theoretical predictions made within the Standard Model. We use the theoretical framework and fundamental constants from \cite{Mohr:2024kco} to determine the energy levels. Correlation coefficients between constants are taken from the corresponding CODATA least-squares adjustment covariance matrix provided by the NIST Physical Constants database \cite{CODATA2022data}, which is not included in the summary publication. Note that, we avoid using the Rydberg frequency $cR_\infty$ due to its strong correlation with the experimental $1\mathrm{S}_{1/2}-2\mathrm S_{1/2}$ transition frequency. Instead, we express it as $cR_\infty=\alpha^2m_ec^2/2h$ and use $\alpha$ and $m_e$ as independent inputs. For the proton rms charge radius $r_p$, we adopt the value extracted from muonic hydrogen measurements \cite{Antognini:2013txn}, implicitly assuming that the scalar under consideration does not couple to muons. This allows us to use the precise $1\mathrm{S}_{1/2}-2\mathrm S_{1/2}$ measurement to constrain the parameter space, since the correlations of the remaining constants with the Rydberg constant are below 0.005 and therefore negligible for our purposes. For antihydrogen, we use the same calculation. Thereby, we restrict the CPT violation to the effects mediated by the new scalar.

The tabulated quantities required for the calculation of the energy levels in \cite{Mohr:2024kco} do not include values for the $6\mathrm P_{1/2}$ and $6\mathrm P_{3/2}$ states. However, they are given in references \cite{Drake:1990zz,Jentschura:2003we,PhysRevA.74.062517,140036,LeBigot:2003zz}. Since the theoretical uncertainty for the $6\mathrm P_j$ states is not provided in \cite{Mohr:2024kco}, we assume $\sigma_\mathrm{th}(2\mathrm S_{1/2})\gg\sigma_\mathrm{th}(6\mathrm P_j)$, consistent with the hierarchy observed for the tabulated $n\mathrm P_j$ states. The resulting uncertainty of the $6\mathrm P_j$ states is therefore subdominant for the $2\mathrm S_{1/2}-6\mathrm P_j$ transitions.

The resulting theoretical transition frequencies and their uncertainties are listed in \autoref{tab:data}. For the uncertainties, we combine the theoretical uncertainties listed in Table XII of reference~\cite{Mohr:2024kco} with those arising from the input constants \cite{CODATA2022data}. In the following analysis, we account for correlations arising from both the theoretical uncertainties and those of the input constants.

\subsection{Fit Procedure}
We define the differences between experimental and theoretical transition frequencies as
\begin{align}
    \delta
    =
    f_\mathrm{exp}-f_\mathrm{theo}.
\end{align}
A consistent sign convention is required in the fit. With this choice, any new physics contributions entering the theoretical prediction must be subtracted from $\delta$. Assuming Gaussian uncertainties, the experimental and theoretical uncertainties are combined by summing the corresponding covariance matrices to obtain the covariance matrix of $\delta$. We then perform a least-squares fit to obtain the confidence regions for the four parameter combinations 
$g_p^+g_e^+$, $g_p^+g_e^-$, $g_p^-g_e^+$ and $g_p^-g_e^-$, following the techniques presented in \cite{Cowan:1998ji,Anderson2003}. The results are shown in \autoref{sec:results}. Further details on the fit procedure and covariance matrix are provided in \autoref{sec:app_stat}.


\section{Results}
\label{sec:results}

In this section, we  apply the  formalism developed in the previous section to obtain limits on the different matter and antimatter coupling combinations. We start with a full fit to a combination of both hydrogen and antihydrogen spectroscopy data. We then demonstrate that we can also obtain limits on the same coupling combinations, including those on antimatter, by using data on hydrogen alone. We highlight that a particularly powerful input is the measurement of the Lamb shift in a beam experiment with significant velocity $\sim 0.01$. Next, highly charged ions contribute valuable information due to larger velocities of both the electron and the nucleons. 

\subsection{Hydrogen and Antihydrogen Spectroscopy}

\begin{figure*}[t!]
\centering
  \includegraphics[width=\textwidth]{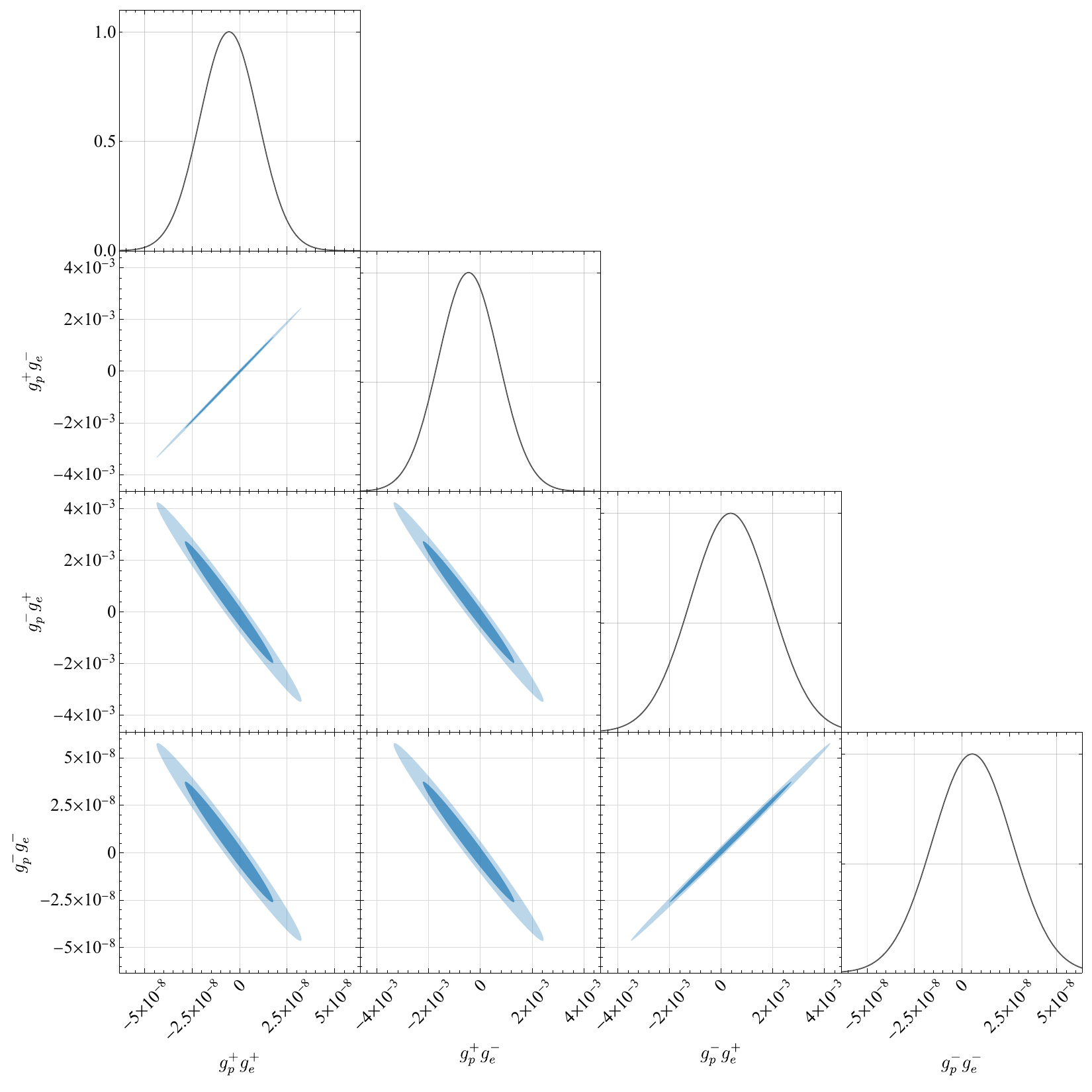}~~
  \caption{Allowed parameter space for non-relativistic matter and antimatter couplings of electrons and protons assuming a scalar mass of $1\:\mathrm{MeV}$. The darker and lighter shaded blue areas correspond to the $1$ and $2\sigma$ contours, respectively. The corresponding Gaussian distributions are shown at the top of each column.}
\label{fig:grid}
\end{figure*}

We use the theoretical and experimental values of the hydrogen and antihydrogen transition frequencies from \autoref{tab:data}, together with the correlation matrix from \autoref{tab:correlation_matrix} given in \autoref{sec:app_stat}, to perform a least-squares fit for the four combinations of the non-relativistic proton and electron couplings as described in \autoref{sec:analysis_framework}. As a benchmark, we assume a scalar mass of $1\:\mathrm{MeV}$ and show the resulting allowed parameter space in \autoref{fig:grid}. The two-dimensional contours are obtained by marginalizing over the remaining two parameters, which amounts to taking the corresponding entries of the full covariance matrix for the parameter products \cite{Anderson2003}.

As expected, the strongest bounds are obtained for the particle-particle coupling $g_p^+g_e^+$. At the same time, the bounds for the antiparticle-antiparticle coupling $g_p^-g_e^-$ are almost equally strong. The corresponding bounds are naturally dominated by hydrogen and antihydrogen data, respectively. The reason is that the total uncertainties from theory and experiment are of the same order of magnitude (cf.~\autoref{tab:data}). This is due to the, in some cases dominating, uncertainties of the input constants used in the theoretical calculation, which affect the hydrogen and antihydrogen data equally.

All fitted parameter combinations are compatible with zero at the $1\sigma$ level. The fit improves from $\chi_\mathrm{min}^2=23.8$ in the absence of scalar interactions to $\chi_\mathrm{min}^2=21.3$ at the best-fit point for $m_\phi=1\:\mathrm{MeV}$, both corresponding to $11$ degrees of freedom. We obtain similar result throughout the considered mass range.

\begin{figure*}[t!]
\centering
  \includegraphics[width=0.9\textwidth]{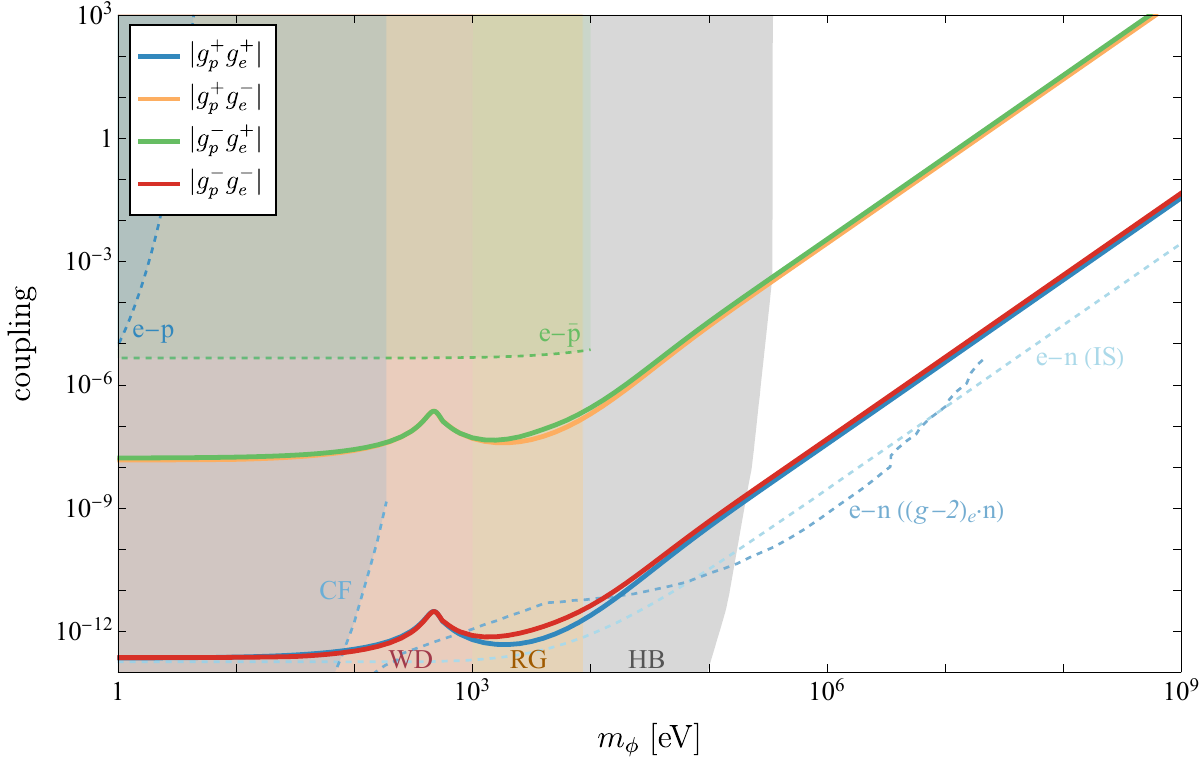}~~
  \caption{Bounds on the absolute values of the products of electron and proton couplings. The solid lines show the 95\% confidence level limits derived in this work, which are obtained from the marginalized one-dimensional distributions. They are compared to laboratory and astrophysical bounds. While the gray, beige and rose areas, corresponding to horizontal branch (HB) stars, red giants (RG) and white dwarfs (WD), respectively, are applicable to all of the considered coupling products, the dashed lines represent laboratory bounds comparable only to the bound of the same color. Electron-proton bounds are given in \cite{Wu:2023yoq}, while the electron-antiproton bounds are from \cite{Ficek2018}. The lower dashed blue lines correspond to electron-neutron bounds from $(g-2)_e$ combined with neutron scattering \cite{Berengut:2017zuo} and hydrogen and deuterium isotope shifts (IS) \cite{Delaunay:2017dku}. We also include Casimir-force (CF) bounds \cite{Klimchitskaya:2025ghy}. See text for details.}
\label{fig:massbounds}
\end{figure*}

In \autoref{fig:massbounds}, we show the mass dependence of the bounds on the absolute values of the possible products of the electron and proton couplings. They are compared to other laboratory and astrophysical bounds, which will be discussed in \autoref{sec:other}. To our knowledge, our bounds provide the strongest direct limits on all coupling products for masses $\gtrsim 0.4\:\mathrm{MeV}$. The ratios of the corresponding bounds on $g_p^+g_e^+$ and $g_p^+g_e^-$, originating from hydrogen measurements, as well as $g_p^-g_e^-$ and $g_p^-g_e^+$, corresponding to antihydrogen, are of order $\mathcal O(10^{-5})$. This is consistent with the expectation that the positron contributions to hydrogen and the electron contributions to antihydrogen are suppressed by $v^2/4\sim\alpha^2/4\sim \mathcal O(10^{-5})$, as discussed in \autoref{sec:energy_correction}.

A crucial role is played by the transitions corresponding to measurements with significant beam velocities, i.e. transitions 3 and 4. In their absence, some of the bounds significantly worsen. The reason is as follows. In order to constrain the four parameter products, the corresponding Fisher matrix needs to have rank 4. It follows from \autoref{eq:Delta_E_H_v} that hydrogen transitions measured at cryogenic temperatures are only sensitive to the combinations $g_p^+g_e^+-\frac{v_\mathrm{CMB}^2}{4}g_p^-g_e^+$ and $g_p^+g_e^--\frac{v_\mathrm{CMB}^2}{4}g_p^-g_e^-$ with $\frac{v_\mathrm{CMB}^2}{4}\sim 10^{-7}$, appearing identically in every such transition. In this case, hydrogen spectroscopy alone cannot separate $g_p^+g_e^+$ from $g_p^-g_e^+$, or $g_p^+g_e^-$ from $g_p^-g_e^-$ as there is always a small admixture of the latter of the two coupling products. Hence, without transitions 3 and 4, hydrogen alone only contributes two linearly independent directions, such that the Fisher matrix has rank 2. Thus, we also need at least two different transitions for antihydrogen to raise the rank of the Fisher matrix to 4. While the experimental precision of the $1\mathrm{S}_{1/2}-2\mathrm S_{1/2}$ transition frequency of antihydrogen is better than the theory uncertainty, the measurements of the other two included antihydrogen transitions are far less precise. Now using that antihydrogen is sensitive to $g_p^-g_e^+-\frac{v_\mathrm{CMB}^2}{4}g_p^+g_e^+$ and $g_p^-g_e^--\frac{v_\mathrm{CMB}^2}{4}g_p^+g_e^-$, i.e. the reverse combinations as can be seen from \autoref{eq:Delta_E_Hbar_c}, it follows that the bounds on the coupling products $g_p^-g_e^+$ and $g_p^-g_e^-$, which are primarily identified with antihydrogen, suffer from the higher experimental uncertainty. Consequently, this low sensitivity propagates to $g_p^+g_e^+$, whose bound is degraded to the level of $v_\mathrm{CMB}^2/4$ times the weak antihydrogen bound on $g_p^-g_e^+$, despite the naive expectation that the exceptional precision of hydrogen measurements will lead to a strong bound on the matter coupling product $g_p^+g_e^+$. This degeneracy is broken by the addition of hydrogen transitions measured at a velocity larger than $v_\mathrm{CMB}$. Crucially, the inclusion of the transitions 3 and 4, performed on atomic beams with $v_\mathrm{exp}\sim 0.01$, lead to an increase of the hydrogen contribution to the rank of the Fisher matrix as the linearly independent combination $g_p^+g_e^+-\frac{v_\mathrm{exp}^2}{4}g_p^-g_e^+$ with $v_\mathrm{exp}^2/4\sim 10^{-5}$ is probed. Following from velocity hierarchy, we have replaced $v_\mathrm{CMB}$ by $v_\mathrm{exp}$ in \autoref{eq:Delta_E_H_v}. Due to the much better precision in comparison to the antihydrogen measurements 14 and 15, this means that the latter two measurements essentially decouple. While they were required before to get to a full-rank Fisher matrix, they now become spectators in the least-squares fit. Hence, the resulting bounds mainly correspond to the unsuppressed matter-matter coupling product in hydrogen, $g_p^+g_e^+$, and a corresponding bound on $g_p^+g_e^-$, suppressed by $v^2/4$ as mentioned above, as well as the unsuppressed antimatter-antimatter coupling product in antihydrogen transitions, $g_p^-g_e^-$, with a bound suppressed by $v^2/4$ on $g_p^-g_e^+$.

Finally, we note that the measurements with non-vanishing beam velocities also reduce the frame dependence. For example, the limits hardly change for an Earth-centred reference frame for the couplings.

\subsection{Constraining Antimatter Couplings with Matter using only Hydrogen Spectroscopy}

\begin{figure*}[t!]
\centering
  \includegraphics[width=0.9\textwidth]{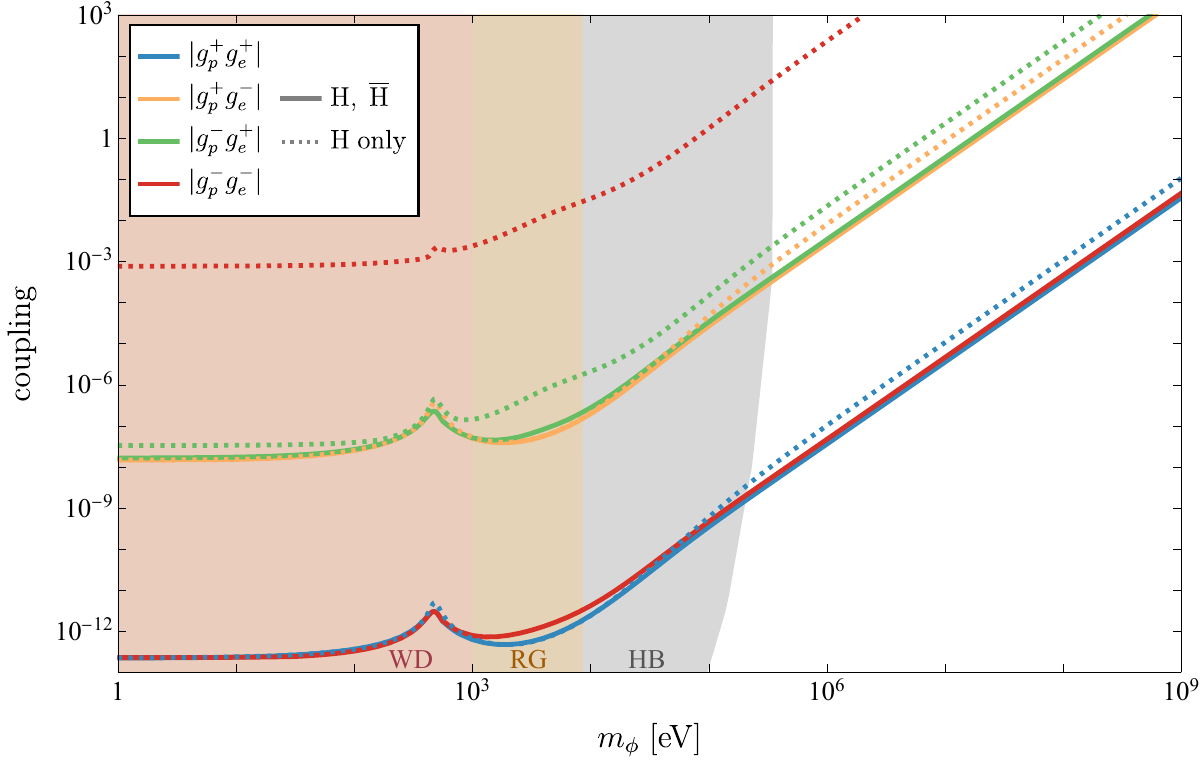}~~
  \caption{Comparison between bounds at 95\% CL originating from hydrogen data alone (dotted) and combined hydrogen and antihydrogen data (solid, as in \autoref{fig:massbounds}) on the absolute values of the products of electron and proton couplings. The shaded regions correspond to astrophysical bounds from horizontal branch (HB) stars, red giants (RG) and white dwarfs (WD) and are applicable to all of the considered coupling products.}
\label{fig:massbound_H}
\end{figure*}

A main result of the present paper is that we can constrain the antimatter couplings by using exclusively matter measurements. To do so, in the following, we only consider the transitions indexed with the numbers 1 to 10, i.e. we restrict the analysis to hydrogen transitions. As argued in the previous section, the rank of the Fisher matrix and hence the covariance matrix can be increased by including measurements of hydrogen transitions performed with moving atoms. With transitions at two different beam velocities that dominate over $v_\mathrm{CMB}$, as is the case for transitions 3 and 4, a full-rank matrix can be obtained, enabling bounds to be placed on all four parameter products. The corresponding bounds are shown as the dotted lines in \autoref{fig:massbound_H} and make the velocity hierarchies following from \autoref{eq:Delta_E_H_v} manifest. The matter-matter coupling product $g_p^+g_e^+$ is unsuppressed and its bound sets the benchmark. Due to the electron-velocity, the bound on $g_p^+g_e^-$ is suppressed by $v^2/4\sim10^{-5}$ as in the previous section. For the bound on $g_p^-g_e^+$, we need to consider $v_\mathrm{exp}^2/4\sim 10^{-5}$ instead of $v_\mathrm{CMB}^2/4$ as the suppression factor compared to the bound on $g_p^+g_e^+$. The different shape of the curve appears as the transitions 3 and 4, which contribute strongly, have a different radial behaviour compared to the transitions that contribute to the previously discussed bounds. Lastly, the antimatter-antimatter coupling product is suppressed by a further factor of $v^2/4\sim 10^{-5}$ corresponding to the electron velocity. As this coupling is now mostly dominated by transitions 3 and 4, the typical bump slightly below $m_\phi\simeq 1\:\mathrm{keV}$ has almost disappeared.

We have thus shown that we can constrain all four parameter products, in particular including the ones containing antimatter couplings, using hydrogen spectroscopy alone. As expected, the antiparticle coupling bounds are not competitive compared to the ones obtained when including antihydrogen spectroscopy.

\subsection{Highly Charged Ion Spectroscopy}

\begin{table}[t]
    \centering
    \begin{tabular}{c c c c c c c c}
    \hline\hline
        Index  & Ion & Transition & Velocity & Theory (eV) & Experiment (eV) & Refs.\\
        \hline
         \rule{0pt}{0.8\normalbaselineskip}1&$^{132}\mathrm{Xe}^{53+}$&$1\mathrm S_{1/2}-2\mathrm P_{3/2}$&$0.31$&$31\:283.948\:(19)$& $31\:284.00\:(287)$&\cite{herdrich2023high}\\
         2&$^{197}\mathrm{Au}^{78+}$&$1\mathrm S_{1/2}-2\mathrm P_{3/2}$&$0.63$&$71\:570.43\:(22)$&$71\:573.11\:(790)$&\cite{beyer1995measurement,Indelicato:2019nij}\\
         3&$^{238}\mathrm{U}^{91+}$&$1\mathrm S_{1/2}-2\mathrm P_{3/2}$&$0.30$&$102\:175.10\:(54)$& $102\:178.12\:(433)$&\cite{gumberidze2005quantum,Indelicato:2019nij} \\
         4&$^{238}\mathrm{U}^{91+}$&$1\mathrm S_{1/2}-2\mathrm P_{1/2}$&$0.67$&$97\:611.95\:(54)$& $97\:605.61\:(1600)$&\cite{beyer1995measurement,Indelicato:2019nij}\\
         \hline\hline
    \end{tabular}
    \caption{Transitions of hydrogen-like highly charged ions included in the analysis. Theoretical transition energies are taken from~\cite{yerokhin2015lamb}, while the experimental transition energies and the corresponding ion beam velocities are taken from the references listed in the \textit{Refs.} column. All energies correspond to the hyperfine centroids.}
    \label{tab:data_ion}
\end{table}

So far, we have considered spectroscopy of hydrogen and antihydrogen as well as hydrogen alone. Due to the high precision of both experiments and theory, we were able to set stringent bounds on the parameter products of the considered model. We now go beyond hydrogen and consider highly charged hydrogen-like ions, i.e. ions with only 1 electron. While the spectroscopy of highly charged ions is much less precise, we gain on several counts. Firstly, the mean distance between electron and nucleus scales as $1/Z$, such that more compact orbitals allow us to probe heavier mediators for atoms with larger $Z$. Secondly, since the electron velocity scales as $Z\alpha$, the non-relativistic positron coupling is less suppressed. Concerning the nucleons, their movement inside the nuclear core is substantial as we show below, such that the products including the non-relativistic antinucleon coupling is far less suppressed as well. This effect is actually surpassed by typically large ion beam velocities, strongly enhancing the sensitivity to antimatter couplings. In the following, we make simple assumptions and approximations to illustrate the impact of these effects and estimate bounds on the coupling products in the considered framework from the spectroscopy of hydrogenic heavy ions. The considered ions and their transition frequencies are given in \autoref{tab:data_ion}.

We estimate the velocity of the nucleons inside the nuclei assuming a three dimensional harmonic oscillator model \cite{goeppert1955elementary,bohr1970nuclear}. As we will see below, the resulting nucleon velocities are moderately relativistic. For simplicity, we nevertheless use a non-relativistic approximation, noting that the resulting corrections are irrelevant as the beam velocities dominate. Using the virial theorem, we relate the kinetic energy to the energy levels of the harmonic oscillator, $\langle T\rangle=\frac 12E_N=\frac 1 2(N+\frac 3 2)\hbar \omega$, where $\hbar\omega=(45A^{-1/3}-25A^{-2/3})\:\mathrm{MeV}$ with $A$ the sum of protons and neutrons in the nuclear core~\cite{Blomqvist:1968zz}. Since the nucleons occupy different oscillator shells, we average the quantity entering the kinetic energy, $\langle N+\frac 32\rangle=\frac1A \sum_i(N_i+\frac 3 2)$, over all occupied states. The oscillator shells are filled from lowest to highest energy, each shell having a capacity $(N+1)(N+2)$ for both protons and neutrons, respectively. We find the averages $\langle N+\frac 32\rangle_{^{132}\mathrm{Xe}}\approx 4.5$, $\langle N+\frac 32\rangle_{^{197}\mathrm{Au}}\approx 5.1$ and $\langle N+\frac 32\rangle_{^{238}\mathrm U}\approx 5.5$ for the ions under consideration. This can be translated to the root-mean-square of the velocity, $v_N^\mathrm{rms}\equiv\sqrt{\langle v_N^2\rangle}=\sqrt{\hbar\omega\langle N+\frac32\rangle/m_N}$, leading to $v_N^\mathrm{rms}\approx 0.2$ for all three of the considered nuclei. Thus, although $\langle N+\frac 32\rangle$ increases with $A$, this is largely compensated by the decrease of $\hbar\omega$ with $A$, leading to nearly identical rms velocities.

To obtain the energy shift due to the presence of the new scalar particle in hydrogenic ions, we can suitably adapt the result from hydrogen from \autoref{eq:Delta_E_H_c}. First, since we consider the transitions in the ionic rest frame and the measurements are performed at beams with large velocities $v_\mathrm{exp}$, a boost from the CMB frame to the ionic rest frame is required, such that $c_\pm^\mathrm{exp}\equiv (\gamma_\mathrm{exp}\pm 1)/2$ replaces $c_\pm$. Second, even in the ion rest frame, the nucleons are not at rest but feature the intrinsic velocity $v_N^\mathrm{rms}$ derived above. Continuing with our rough non-relativistic approximation, we estimate the total effect of the beam and the intrinsic velocities by adding them in quadrature, $v_\mathrm{exp}^2\rightarrow v_\mathrm{exp}^2+(v_N^\mathrm{rms})^2$. These, in turn, enter $c_\pm^\mathrm{exp}$ via $\gamma_\mathrm{exp}$. We assume the scalar to couple equally to protons and neutrons, leading to corresponding enhancement factors. 

For the electron, we use the hydrogenic spinor from \autoref{eq:spinor_hydrogen}. This assumes a point-like nucleus and we note that the finite size as well as possible deformations of the larger nuclei may lead to sizable corrections. Using this, we find the energy shift
\begin{align}
    \Delta E_\mathrm{Ion}^Q
    &=
    -A\frac{c_+^\mathrm{exp}g_N^+-c_-^\mathrm{exp}g_N^-}{4\pi}\left[g_e^+\left(c_+^\mathrm{exp}I_G+c_-^\mathrm{exp}I_F\right)-g_e^-\left(c_-^\mathrm{exp}I_G+c_+^\mathrm{exp}I_F\right)\right]\label{eq:Delta_E_Ion_c},
\end{align}
where we have replaced $g_p^\pm$ by $g_N^\pm$ to account for equal nucleon coupling.

\begin{figure*}[t!]
\centering
  \includegraphics[width=0.9\textwidth]{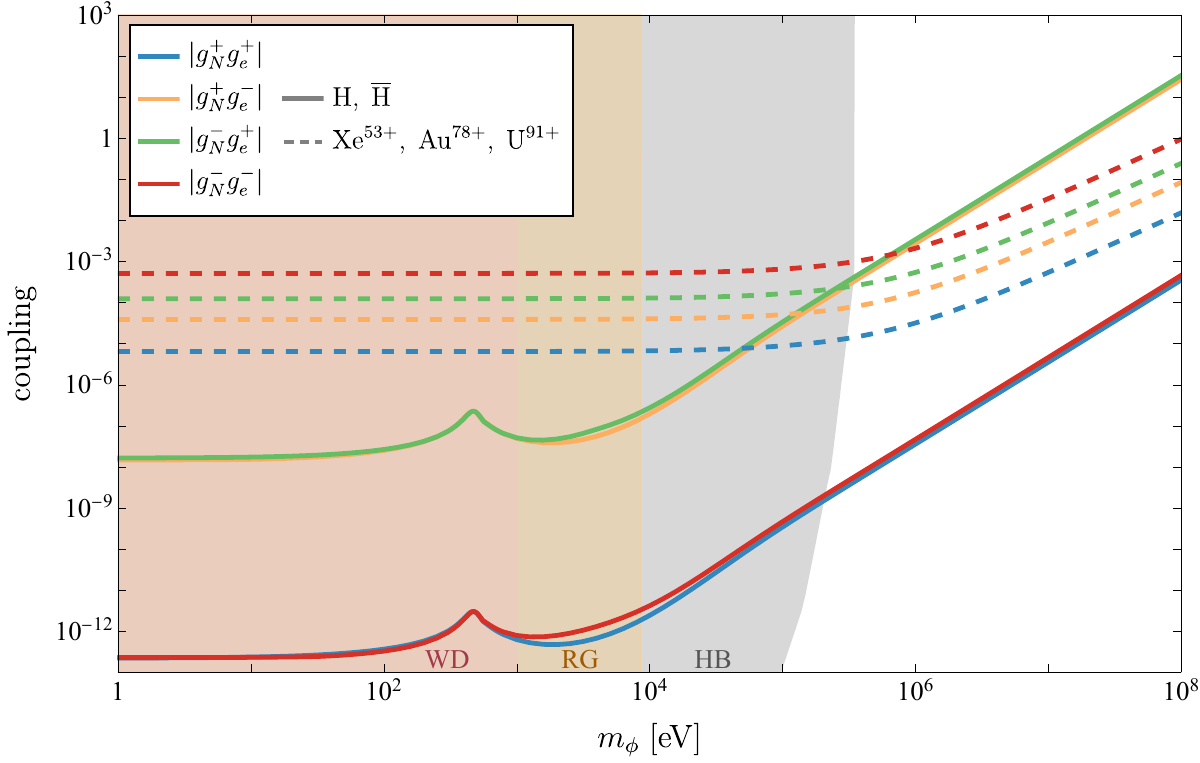}~~
  \caption{Comparison between bounds at 95\% CL originating from highly charged ion data (dashed) and combined hydrogen and antihydrogen data (solid, as in \autoref{fig:massbounds}) on the absolute values of the products of electron and nucleon couplings. As in previous figures, the shaded regions correspond to astrophysical bounds from horizontal branch (HB) stars, red giants (RG) and white dwarfs (WD). They are applicable to all of the considered coupling products.}
\label{fig:ion_bounds}
\end{figure*}

As for the hydrogen case, we perform a least-squares fit to the transitions shown in \autoref{tab:data_ion}. The theory values are derived from \cite{yerokhin2015lamb}. The resulting bounds on the absolute values of the parameter products are shown in \autoref{fig:ion_bounds}. While the bounds are much weaker in the low-mass regime, the bounds on $g_N^+g_e^-$ and $g_N^-g_e^+$ exceed the ones obtained from the combined hydrogen and antihydrorgen fit. This can be understood as follows. Due to the large velocities of both the nucleons and the electron with respect to the CMB rest frame, the accessibility to the antimatter couplings is strongly enhanced. Hence, the four lines lie much closer together and differ by less than an order of magnitude. Because of the much smaller distance between nucleus and electron, the drop in sensitivity sets in at much larger masses compared to hydrogen. The bounds also weaken more slowly because relativistic effects at larger values of $Z\alpha$ increase the probability of the electron to be near the nucleus.\footnote{For example for the $1\mathrm{S}_{1/2}$ state the wave function squared behaves as $\sim r^{2(\sqrt{(1-(Z\alpha)^2}-1)}$, resulting in $\sim r^{-0.52}$ for the case of uranium.} 
As this effect is sensitive to finite nuclear size, we note that we expect corrections from this to become important for masses $m_\phi\gtrsim 30\:\mathrm{MeV}$.

\section{Other Bounds}\label{sec:other}
For comparison, we provide estimates for some limits that can be obtained by other means. 

\subsection{Laboratory Bounds}
We compare our bounds to limits from experiments that probe specific combinations of particle and antiparticle couplings in the non-relativistic limit (see \cite{Cong:2024qly} for a recent review). We stress that all of the following laboratory bounds were translated very naively, without a careful treatment of the relevant velocities. Hence, they should only be taken as indicative of the rough order of magnitude that can be achieved. 

In \cite{Wu:2023yoq}, a spin-mechanical quantum chip is used to obtain bounds on the scalar-scalar coupling between electrons and nucleons. In our case, this translates to the coupling $g_p^+g_e^+$ with the bound indicated by the dashed blue line and area in the top left of \autoref{fig:massbounds}. Constraints on antiproton and electron couplings are obtained from spectroscopy of antiprotonic helium \cite{Ficek2018}. These results constrain $g_p^-g_e^+$, represented by the dashed green line and area in \autoref{fig:massbounds}. For completeness, we also include limits on electron-neutron couplings, which are transferable to the electron-proton couplings $g_p^+g_e^+$ under the assumption that nucleons couple to the scalar with equal strength. The darker blue dashed line corresponds to $(g-2)_e$ bounds combined with neutron scattering data \cite{Berengut:2017zuo}, while the light blue dashed line refers to isotope shift spectroscopy of hydrogen and deuterium \cite{Delaunay:2017dku}. We derive Casimir-force bounds assuming equal coupling to protons and neutrons. For couplings of the same sign, the weakest constraint on the product $g_p^+g_e^+$ is obtained when the electron and nucleon contributions to the macroscopic scalar charge are of comparable size, i.e. $g_p^+=\frac ZAg_e^+$. Using the limits of \cite{Klimchitskaya:2025ghy}, the corresponding bound is shown by the shaded region on the left of \autoref{fig:massbounds}. We note, however, that in principle, cancellation between electron and nucleon couplings is possible for opposite-sign couplings, such that a bound on the product of the couplings may not exist.

\subsection{Astrophysical Bounds}
\label{sec:astro_bounds}
Many astrophysical systems are sensitive to contributions from new physics due to the additional energy loss induced by the production of light particles (see~\cite{Raffelt:1996wa} for an overview). In the following, we discuss bounds from astrophysical environments such as red giants (RGs), horizontal branch (HB) stars and white dwarfs (WDs), adopting suitable order-of-magnitude approximations for our model. The dominant process usually is bremsstrahlung, e.g. $e^-+Z\rightarrow e^-+Z+\phi$, where the scattering of an electron off an ion with charge $Ze$ leads to the emission of a scalar boson $\phi$. Since the astrophysical systems under consideration consist of particles rather than antiparticles, access to the non-relativistic antiparticle couplings $g_F^-$ requires relativistic corrections, analogously to the atomic case. 

The emissivity per unit volume is given by (cf., e.g.~\cite{Raffelt:1996wa,Carenza:2025jwn}) 
\begin{align}
\label{eq:emissivity}
    Q_\phi
    &=
    \int\frac{2\mathrm d^3p_i}{(2\pi)^32E_i}\int\frac{2\mathrm d^3p_f}{(2\pi)^32E_f}\int\frac{\mathrm d^3p_\phi}{(2\pi)^32\omega_\phi}\omega_\phi(2\pi)\delta(E_i-E_f-\omega)f_i(1-f_f)\vert\overline{\mathcal M}\vert^2,
\end{align}
where $f_{i,f}$ are the corresponding Fermi-Dirac distribution functions at the initial and final energies $E_i$ and $E_f$ of the particle emitting the scalar, $\omega_\phi$ is the energy carried by the emitted scalar and $\vert\overline{\mathcal M}\vert^2$ is the spin-averaged amplitude squared that carries the information about the interaction. 

We roughly estimate
\begin{align}
\label{eq:matrixsimplified}
    \vert\overline{\mathcal M}\vert^2
    &\propto
    \vert\overline{\mathcal B}\vert^2(\eta)
    \simeq
    \frac{(E_f+m)(E_i+m)}{4m^2}\left[g_F^+-\eta\frac{g_F^-\vert\boldsymbol{p}_f\vert\vert\boldsymbol{p}_i\vert}{(E_f+m)(E_i+m)}\right]^2
\end{align}
with $\vert\overline{\mathcal B}\vert^2$ the spin-averaged and squared bilinear of the vertex emitting the scalar particle. We introduced a parameter $\eta$, which is an $\mathcal O(1)$ number subsuming uncertainties in the approximation. Importantly, this structure features the $v^2$-suppression of the non-relativistic antimatter coupling. As in the expression for the emissivity from \autoref{eq:emissivity}, we integrate over the momenta to capture the thermal distribution of the relevant particles,
\begin{align}
     Q_\phi
     \sim
     \langle\vert\overline{\mathcal B}\vert^2(\eta)\rangle
     &=
     \int\frac{4\mathrm dp\,p^2}{2\pi^24E^2}4m^2f(p)(1-f(p))\vert\overline{\mathcal B}\vert^2(\eta),
\end{align}
where we have discarded the kinematic factor from \autoref{eq:matrixsimplified} as well as the Coulomb propagator in the integral. When taking the relevant ratio to obtain our bounds below, we do not expect a large impact from these factors.

There are numerous references discussing bounds on the coupling $g_F$ from astrophysical sources in a simple framework assuming an additional scalar particle with standard Yukawa couplings $g_F\phi\bar\psi_F\psi_F$, including \cite{Raffelt:1996wa,Hardy2017,Bottaro:2023gep,Hardy:2024gwy,Fiorillo:2025zzx,Dev:2020jkh}. We generalize these bounds to our scenario. In the usual scalar case, one obtains the bilinear shown in \autoref{eq:matrixsimplified} with $I_0^F$ set to zero, which we call $\mathcal B_Y=g_F\tilde{\mathcal B}_Y$, where the subscript $Y$ indicates the standard Yukawa case and $\tilde{\mathcal B}_Y(\eta)$ is independent of $g_F$. With the same assumptions as in the previous case, we find the scaling
\begin{align}
    Q_\phi^Y
    \sim
    \langle\vert\overline{\mathcal{B}_Y}\vert^2(\eta)\rangle
    =
    g_F^2\langle\vert\overline{\tilde{\mathcal B}_Y}\vert^2(\eta)\rangle
    =
    g_F^2\int\frac{4\mathrm dp\,p^2}{2\pi^24E^2}4m^2f(p)(1-f(p))\vert\overline{\tilde{\mathcal B}_Y}\vert^2(\eta).
\end{align}
Since the observational bound $Q_\phi<Q_\mathrm{obs}$ constrains the total emissivity independently of the underlying model, and the same bound yields $g_F<g_F^\mathrm{bound}$ when applied to the vanilla Yukawa case, taking the ratio directly constrains the bilinear structure,
\begin{align}
\label{eq:astro_bound}
    \frac{\langle\vert\overline{\mathcal B}\vert^2(\eta)\rangle}{\langle\vert\overline{\tilde{\mathcal B}_Y}\vert^2(\eta)\rangle}
    =
    \frac{\int_0^\infty\mathrm dp\,\frac{p^2}{E^2}f(p)(1-f(p))\vert\overline{\mathcal B}\vert^2(\eta)}{\int_0^\infty\mathrm dp\,\frac{p^2}{E^2}f(p)(1-f(p))\vert\overline{\tilde{\mathcal B}_Y}\vert^2(\eta)}
    <
    \left(g_F^\mathrm{bound}\right)^2.
\end{align}
After determining the chemical potential for a given astrophysical environment, Lagrange multiplier methods can be applied to obtain bounds on $\vert g_F^+\vert$ and $\vert g_F^-\vert$ with a dependence on $\eta$.

While this approximation does not capture the full kinematic structure of the bremsstrahlung process, it is sufficient for the order-of-magnitude estimates considered in the following. As we will see, the resulting astrophysical bounds exceed those obtained from spectroscopy by several orders of magnitude, making the approximation suitable for our purposes.\footnote{Finally, we note that astrophysical bounds are typically derived assuming a coupling to only one species at a time, with all other couplings set to zero. When couplings to both electrons and nucleons are present simultaneously, interference terms between diagrams with scalar emission from either particle arise. We neglect these interference terms here and note that a complete treatment would require their explicit evaluation.} A more thorough consideration will be provided in~\cite{Lucenteprep}.

\subsubsection*{Bounds from Red Giants}
Since our approximate expression given above is based on bremsstrahlung as the energy loss mechanism, we use the limits from~\cite{Bottaro:2023gep} for our rough translation, although a careful investigation of the resonant plasmon conversion considered in~\cite{Hardy2017} may lead to stronger limits.
The limits from~\cite{Bottaro:2023gep} are
\begin{align}
    g_F^\mathrm{bound}\lesssim\begin{cases}
        4\times10^{-15}&\mathrm{electrons},\\
        7\times 10^{-12}&\mathrm{nucleons},
    \end{cases}
\end{align}
where the latter bound was obtained assuming that $\phi$ couples equally to protons and neutrons. The bounds were derived with the stellar parameters $T_\mathrm{RG}\sim 8.6\:\mathrm{keV}$ and $\rho_\mathrm{RG}\sim 10^6\:\mathrm{g}\:\mathrm{cm}^{-3}$ 
and a nuclear composition of mostly $^4\mathrm{He}$ cores.\footnote{\label{foot:astro_He}The intrinsic velocities of the nucleons inside the nucleus are much larger than the thermal velocity of the nucleus. In principle, this leads to a relatively large effective coupling of the helium nucleus even in the case of a pure antinucleon coupling. However, using this there still exists a combination of the nucleon and antinucleon couplings such that the helium nucleus does not couple to the scalar. This would lead to a degenerate direction. This degeneracy is broken by thermal velocities. For our rough estimate we therefore neglect the intrinsic velocity. Yet, we note that a more careful treatment will likely change the antinucleon limits by several orders of magnitude, possibly strengthening them.}
Our rough translation then gives, 
\begin{align}
\begin{split}
\label{eq:RG_bounds}
    \vert g_e^+\vert&\lesssim 10\:g_e^\mathrm{bound}\sim 4\times 10^{-14},\\
    \vert g_e^-\vert&\lesssim\frac{80}{\eta}g_e^\mathrm{bound}\sim 3\times 10^{-13},\\
    \vert g_N^+\vert&\lesssim 1.6\:g_N^\mathrm{bound}\sim 1\times 10^{-11},\\
    \vert g_N^-\vert&\lesssim\frac{7.1\times 10^5}{\eta}g_N^\mathrm{bound}\sim 5\times 10^{-6},
\end{split}
\end{align}
where we assumed $\eta=1$ as an estimate in the respective last steps. The approximate dependence on $\eta$ is only valid for $\eta\lesssim\mathcal O(1)$, particularly for the electron bounds, with correction terms playing an increasingly important role as $\eta$ increases. As we can see from \autoref{eq:matrixsimplified}  
the parameter $\eta$ controls the antiparticle contribution and is accordingly reflected in the above formulas. We note that the Fermi-Dirac distribution in \autoref{eq:astro_bound} should technically be replaced by a Bose-Einstein distribution for the helium nuclei in a stellar core. However, as the nuclei are non-degenerate, this has no numerical effect.

For the RG core temperature $T_\mathrm{RG}\sim 8.6\:\mathrm{keV}$, the thermal velocities of both nuclei and electrons dominate over the typical motion of stars inside the Milky Way with respect to the CMB rest frame. We therefore neglect the latter.

For electron-ion bremsstrahlung in RGs, a full calculation for the electron couplings within the framework of the considered model was actually performed in~\cite{Carenza:2025jwn} in a one-zone model with density $\rho=2.5\times 10^{5}\:\mathrm g\:\mathrm{cm}^{-3}$ and temperature $T=8.6\:\mathrm{keV}$. For $m_\phi\lesssim T$ and setting all other considered parameters to zero, the constraint from~\cite{Carenza:2025jwn} is,
\begin{align}
    23g_e^2+59g_eI_0^e+56{I_0^e}^2<2.2\times 10^{-28}\equiv b,
\end{align}
where the bound $b$ is based on limits on the axion emissivity and the coupling combination is derived by numerically calculating the emission rate including the respective couplings. To find the corresponding non-relativistic electron and positron couplings, we convert the bound and find
\begin{align}
\begin{split}
    \vert g_e^+\vert&<\sqrt{\frac{80}{1671}b}\simeq 3.3\times 10^{-15},\\
    \vert g_e^-\vert&<\sqrt{\frac{184}{557}b}\simeq 8.5\times 10^{-15}.
\end{split}
\end{align}
The relatively small difference between the matter and the antimatter bound is likely a consequence of the partial degeneracy. 

Comparing to \autoref{eq:RG_bounds}, we see that the full calculation gives a stronger limit, but is also not too far away, strengthening the confidence of our results within a margin of a couple orders of magnitude.

The constraints on the combinations of the electron and nucleon coupling from red giants are shown as the beige area in \autoref{fig:massbounds}, \autoref{fig:massbound_H} and \autoref{fig:ion_bounds}.
The limit shown in the figure corresponds to the product of antiparticle bounds, being the weakest out of the four possible combinations.

\subsubsection*{Bounds from Horizontal Branch Stars}
Horizontal branch (HB) star cores undergo a helium-burning phase, during which additional energy-loss processes from new physics lead to contraction of the core, such that the duration of helium-burning shortens due to increased heat. We can follow the same strategy as in the case of red giants.

For the parameters $T_\mathrm{HB}\sim 10\:\mathrm{keV}$ and $\rho_\mathrm{HB}\sim 10^4\:\mathrm g\:\mathrm{cm}^{-3}$, bounds from $^4\mathrm{He}$ continuum bremsstrahlung (see \autoref{foot:astro_He}) in the low-mass regime $m_\phi\ll T_\mathrm{HB}$ are given by \cite{Hardy2017}
\begin{align}
    g_F^\mathrm{bound}\lesssim\begin{cases}
        1.4\times10^{-14}&\mathrm{electrons},\\
        2.5\times 10^{-11}&\mathrm{nucleons},
    \end{cases}
\end{align}
again assuming equal nucleon couplings. In our framework, these bounds translate to
\begin{align}
\begin{split}
\label{eq:HB_bounds}
    \vert g_e^+\vert&\lesssim 1.7\:g_e^\mathrm{bound}\sim 2\times 10^{-14},\\
    \vert g_e^-\vert&\lesssim\frac{80}{\eta}g_e^\mathrm{bound}\sim 1\times 10^{-12},\\
    \vert g_N^+\vert&\lesssim 1.6\:g_N^\mathrm{bound}\sim 4\times 10^{-11},\\
    \vert g_N^-\vert&\lesssim\frac{6.1\times 10^5}{\eta}g_N^\mathrm{bound}\sim 2\times 10^{-5},
\end{split}
\end{align}
where we take $\eta=1$ as a rough estimate and again note that the given dependence on $\eta$ is only valid for $\eta\lesssim \mathcal O(1)$ with corrections for larger $\eta$. For nucleons, we observe a similar functional dependency as for RGs, resulting from the non-relativistic and non-degenerate nature of the helium nuclei in both environments. In the electronic case, the antiparticle bounds are more strongly suppressed compared to the RG case due to the smaller density and hence the smaller degeneracy, limiting the enhancement of the antiparticle contribution. At core temperatures $T_\mathrm{HB}\sim 10\:\mathrm{keV}$, the thermal velocity of the helium nuclei is similar to the typical velocity relative to the CMB, whereas the electron velocities are much larger. Hence, we neglect the relative velocity to the CMB.

The treatment of the continuum production is done more carefully in \cite{Hardy2017}, such that bounds extend well beyond $m_\phi\sim T_\mathrm{HB}$. We use these mass-dependent bounds, multiplied by the appropriate factors, assuming $\eta=1$ throughout. 

Results are shown in gray in Figures~\ref{fig:massbounds}, \ref{fig:massbound_H} and~\ref{fig:ion_bounds}. As before, we display only the weakest combination.

\subsubsection*{Bounds from White Dwarfs}
Due to their large density, white dwarfs (WDs) provide an interesting environment for scalar emission bounds. The cooling is well-controlled as nuclear fuel is no longer burnt. Ref. \cite{Bottaro:2023gep} finds the bremsstrahlung limits
\begin{align}
    g_F^\mathrm{bound}\lesssim\begin{cases}
        3.9\times10^{-16}&\mathrm{electrons},\\
        6.5\times 10^{-13}&\mathrm{nucleons},
    \end{cases}
\end{align}
assuming equal nucleon coupling and an
equal mixture of carbon and oxygen.\footnote{In this case, the typical velocities of the nuclei are actually smaller than $v_\mathrm{CMB}$. However, we ignore this effect and focus on the thermal velocities. This gives a very rough but conservative limit as these are smaller than both $v_\mathrm{CMB}$ and the intrinsic nucleon velocities. As an aside we note that the slightly different intrinsic nucleon velocities in carbon and oxygen may allow for a breaking of potential degeneracies, but this requires a careful treatment of them.} While their calculation uses a non-trivial density profile, the results are consistent with a one-zone model with constant density $\rho_\mathrm{WD}=1.3\times 10^{6}\:\mathrm g\:\mathrm{cm}^{-3}$. The bounds extend only up to the typical temperature of WDs, i.e. $m_\phi\lesssim 1\:\mathrm{keV}$. With these parameters, we find the bounds
\begin{align}
\begin{split}
\label{eq:WD_bounds}
    \vert g_e^+\vert&\lesssim 100\:g_e^\mathrm{bound}\sim 4\times 10^{-14},\\
    \vert g_e^-\vert&\lesssim\frac{700}{\eta}g_e^\mathrm{bound}\sim 3\times 10^{-13},\\
    \vert g_N^+\vert&\lesssim 1.6\:g_N^\mathrm{bound}\sim 1\times 10^{-12},\\
    \vert g_N^-\vert&\lesssim\frac{1.9\times 10^7}{\eta}g_N^\mathrm{bound}\sim 1\times 10^{-5}.
\end{split}
\end{align}
The effects of degeneracy and relativistic velocities are clearly visible for the electron bounds. The corresponding parameter space ellipse is strongly tilted, leading to a weakening of the particle-like coupling bound by $10^2$ with respect to the original bound $g_e^\mathrm{bound}$. The nucleon antiparticle bounds are more strongly suppressed than in the previously discussed stellar environments due to the lower temperature and hence smaller velocities.

Again, the resulting weakest limit estimates are shown in rose in Figures~\ref{fig:massbounds}, \ref{fig:massbound_H} and~\ref{fig:ion_bounds}.

\subsubsection*{Bounds from Neutron Stars and Supernovae}
Neutron stars and SN~1987A (see~\cite{Hardy:2024gwy,Fiorillo:2025zzx} for some recent incarnations) provide constraints on scalar couplings to nucleons and electrons that are complementary to the bounds derived from RGs, HB stars and WDs. 
In particular, they usually reach higher masses. However, they require a more careful treatment, including effects of high density and re-scattering. Notably, those from supernovae usually also feature a maximal coupling strength above which they do not apply. This is particularly relevant as the spectroscopic bounds are on combinations of electron and nucleon couplings which requires 1) having constraints on both couplings in the desired mass range and 2) that the constraints cannot be avoided by any combination of electron and nucleon couplings that give the desired result (see~\cite{quintprep} for a discussion of this issue). We therefore leave even an estimate of the constraints on these couplings to future work.

\section{Conclusions}\label{sec:conclusions}
It is an intriguing possibility that a dark sector could distinguish between matter and antimatter. In the Standard Model sector, this could then express itself in a new, feebly interacting light particle that couples to matter and antimatter differently, or in the most extreme case, to antimatter only.
The latter case directly motivates experiments with antimatter. However, as a distinction between matter and antimatter requires the breaking of Lorentz symmetry, whether we have a ``pure'' antimatter coupling is a frame- and velocity-dependent statement. If particles move with respect to the frame in which the coupling is only to antiparticles, they indeed acquire a (velocity suppressed) coupling even in the supposedly pure antimatter coupling case. This can then be used to {\emph{constrain antimatter couplings with matter experiments}}. 

In this paper, we have demonstrated this ability and put it into practice using spectroscopy data from simple one electron atomic and ionic systems to obtain stringent limits on both matter and antimatter couplings. Using current experimental data, adding measurements on antihydrogen enhances the sensitivity for low and moderate masses. At higher ($\gtrsim {\rm MeV}$ masses results from matter-only heavy highly charged ions provide the best limits for some coupling combinations. In general, the spectroscopic limits surpass simple estimates\footnote{A more careful study of astrophysical limits, including a wider range of systems, is planned for forthcoming work~\cite{Lucenteprep}.} of astrophysical constraints for masses $\gtrsim {\rm{few}}\times100\,{\rm{keV}}$.

Having the ability to test antimatter couplings in matter based experiments opens a new powerful handle to test a purely antimatter coupled dark sector, providing stimulating complementarity to experiments with antimatter. It also suggests new avenues to explore with matter experiments, such as performing high-precision spectroscopy on relatively fast beams of atoms/ions, as this significantly enhances the sensitivity to antimatter couplings.

\section*{Acknowledgments}
We thank Cedric de Jonge, Yevgeny Stadnik and Maurice Brezavsek for helpful discussions. This work is funded by the Deutsche Forschungsgemeinschaft (DFG, German Research Foundation) under the Collaborative Research Centre SFB 1225 - 273811115 (ISOQUANT).


\appendix

\renewcommand{\sectionautorefname}{Appendix}
\renewcommand{\subsectionautorefname}{Appendix}

\section{Scalar-Mediated Lorentz-Violating Potential}
\label{sec:app_potential}
We derive the potential for a scalar-mediated potential with scalar and time-like vector couplings, with the latter implying Lorentz violation. This generic procedure has been performed in \cite{Altschul:2012xu,Carenza:2025jwn} (see also~\cite{Moody:1984ba}), and we adapt it to our purposes. Crucially, we consider both particles and antiparticles and keep some vertices relativistic. 

\begin{figure*}[t]
    \centering
    \includegraphics[width=0.75\textwidth]{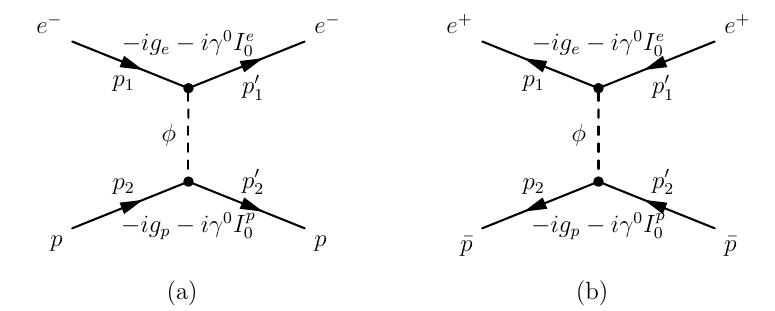}~~
    \caption{Diagrams relevant for a scalar-boson-mediated Lorentz-violating potential with scalar and timelike vector couplings for both particles and antiparticles.}
\label{fig:Feynman_tchannel}
\end{figure*}

We begin with particle-particle scattering and, relating to the nomenclature in \subautoref{sec:potential}, identify $a=p$ as the proton and $b=e$ as the electron. The corresponding $t$-channel amplitude, shown in \autoref{fig:Feynman_tchannel}(a), is given by
\begin{align}
    i\mathcal M
    &=
    \bar u_{s_1'}(\boldsymbol p_1')(-ig_e-i\gamma^0I_0^e)u_{s_1}(\boldsymbol p_1)\frac{i}{q^2-m_\phi^2} \bar u_{s_2'}(\boldsymbol p_2')(-ig_p-i\gamma^0I_0^p)u_{s_2}(\boldsymbol p_2)
\end{align}
with momentum transfer $q=p_1-p_1'=p_2'-p_2$, while the $s_i$ and $s_i'$ correspond to the spin states of the fermions. We now assume the proton to be non-relativistic, i.e. $\vert \boldsymbol{p}_2\vert,\vert\boldsymbol{p}_2'\vert\ll m_p$. We can thus approximate
\begin{align}
    u_s(\boldsymbol p_2)=\sqrt{\frac{E_2+m_2}{2m_2}}\begin{pmatrix}
        \xi_s\\\frac{\boldsymbol{\sigma}\cdot\boldsymbol{p}_2}{E_2+m_2}\xi_s
    \end{pmatrix}
    \simeq
    \begin{pmatrix}
        \xi_s\\\frac{\boldsymbol{\sigma}\cdot\boldsymbol{p}_2}{2m_2}\xi_s
    \end{pmatrix}+\mathcal{O}(\boldsymbol{p}_2^2),
\end{align}
where we have chosen the normalization condition $\bar u_{s'}(\boldsymbol p)u_s(\boldsymbol p)=\delta_{ss'}$, $\xi_s$ is a two-spinor and $\sigma_i$ denote the Pauli matrices. In the non-relativistic limit, the Dirac bilinears of interest are trivially given by
\begin{align}
    \bar u_{s'}(\boldsymbol p')u_s(\boldsymbol p)=\delta_{ss'},\quad u_{s'}^\dagger(\boldsymbol p')u_s(\boldsymbol p)=\delta_{ss'}.
\end{align}
Thus, the amplitude in the semi-relativistic and quasi-static limit ($({q^0})^2\ll \vert\boldsymbol q\vert^2$) becomes
\begin{align}
    i\mathcal M
    =
    \frac{i}{\vert\boldsymbol q\vert^2+m_\phi^2}(g_p+I_0^p)\bar u_{s_1'}(\boldsymbol p_1')(g_e+\gamma^0I_0^e)u_{s_1}(\boldsymbol p_1).
\end{align}
Identifying the electronic initial and final states as the states scattering off the potential generated by the proton, we can write the quantum mechanical amplitude in the Born approximation sense as
\begin{align}
    \mathcal{M}_{\mathrm{QM}}
    =
    -\langle \boldsymbol{p}_1'\vert V\vert\boldsymbol{p}_1\rangle
    =
    \frac{1}{\vert\boldsymbol q\vert^2+m_\phi^2}(g_p+I_0^p)u_{s_1'}^\dagger(\boldsymbol p_1')(g_e\gamma^0+I_0^e)u_{s_1}(\boldsymbol p_1).
\end{align}
The potential is thus given by the Fourier transform of the amplitude,
\begin{align}
    V^{pe}(r)
    &=
    -\int\frac{\mathrm d^3q}{(2\pi)^3}e^{i\boldsymbol{q}\cdot\boldsymbol r}\frac{1}{\vert\boldsymbol q\vert^2+m_\phi^2}(g_p+I_0^p)(g_e\gamma^0+I_0^e)\nonumber\\
    &=
    -(g_p+I_0^p)(g_e\gamma^0+I_0^e)\frac{e^{-m_\phi r}}{4\pi r}
    =
    (g_p+I_0^p)(g_e\gamma^0+I_0^e)f(r),
\end{align}
where $f(r)=-e^{-m_\phi r}/(4\pi r)$ is the Yukawa potential function. In order to obtain the interaction energy of an electron located at distance $r$ from the proton, the potential needs to be contracted with a suitable electron spinor. 

For antiparticles, the above considerations require some modifications. The corresponding Feynman diagram is shown in \autoref{fig:Feynman_tchannel}(b). 
We use the relation $v_s(\boldsymbol p)=\gamma_5u_{-s}(\boldsymbol p)$ to rewrite the matrix element as
\begin{align}
    i\mathcal M
    &=
    \bar v_{s_1}(\boldsymbol p_1)(-ig_e-i\gamma^0I_0^e)v_{s_1'}(\boldsymbol p_1')\frac{i}{q^2-m_\phi^2} \bar v_{s_2}(\boldsymbol p_2)(-ig_p-i\gamma^0I_0^p)v_{s_2'}(\boldsymbol p_2')\nonumber\\
    &=
    -\bar v_{s_1}(\boldsymbol p_1)(-ig_e-i\gamma^0I_0^e)v_{s_1'}(\boldsymbol p_1')\frac{i}{q^2-m_\phi^2} \bar u_{-s_2}(\boldsymbol p_2)\gamma_5(-ig_p-i\gamma^0I_0^p)\gamma_5u_{-s_2'}(\boldsymbol p_2'),
\end{align}
where the sign originates from commuting $\gamma^0$ and $\gamma_5$ in the second bilinear. Note that we only replaced the antiproton spinors as we will use positron spinors in calculations below and that the two minus signs originating from the fields' anticommutation cancel \cite{Altschul:2012xu}. We use the anticommuting nature of $\gamma^0$ and $\gamma_5$ to eliminate the $\gamma_5$ matrices,
\begin{align}
    i\mathcal M
    &=
    -\bar v_{s_1}(\boldsymbol p_1)(-ig_e-i\gamma^0I_0^e)v_{s_1'}(\boldsymbol p_1')\frac{i}{q^2-m_\phi^2} \bar u_{-s_2}(\boldsymbol p_2)(-ig_p+i\gamma^0I_0^p)u_{-s_2'}(\boldsymbol p_2'),
\end{align}
leading to a crucial relative sign flip in the antiproton coupling. Noting that the above non-relativistic relations for the bilinears still hold, we find in the semi-relativistic and quasi-static limit,
\begin{align}
    i\mathcal M
    =
    -\frac{i}{\vert\boldsymbol q\vert^2+m_\phi^2}(g_p-I_0^p)\bar v_{s_1}(\boldsymbol p_1)(g_e+\gamma^0I_0^e)v_{s_1'}(\boldsymbol p_1').
\end{align}
If we now move to the quantum mechanical amplitude relevant for the connection to the interaction potential, we note that the momentum and spin labels retain their positions exactly as in the QFT amplitude since ordering has already been fixed by the Wick contraction, such that
\begin{align}
    \mathcal{M}_{\mathrm{QM}}
    =
    -\langle \boldsymbol{p_1'}\vert V\vert\boldsymbol{p_1}\rangle
    =
    -\frac{1}{\vert\boldsymbol q\vert^2+m_\phi^2}(g_p-I_0^p)v_{s_1}^\dagger(\boldsymbol p_1)(g_e\gamma^0+I_0^e)v_{s_1'}(\boldsymbol p_1').
\end{align}
Fourier transforming as above, we find,
\begin{align}
    V_{e^+\bar p}(r)
    &=
    \int\frac{\mathrm d^3q}{(2\pi)^3}e^{i\boldsymbol{q}\cdot\boldsymbol r}\frac{1}{\vert\boldsymbol q\vert^2+m_\phi^2}(g_p-I_0^p)(g_e\gamma^0+I_0^e)\nonumber\\
    &=
    -(g_p-I_0^p)(g_e\gamma^0+I_0^e)f(r),
\end{align}
with the contraction with a suitable positron spinor implied to obtain the interaction energy.

While the two derived potentials between two particles or between two antiparticles are the ones relevant to this work, we note that they take the same form when replacing the electron with a positron and vice versa up to a sign, which leads to attractive scalar-scalar coupling in all cases. This requires some straight-forward modifications in the derivation.

\subsection{Full Potential in the Laboratory Frame}
\label{sec:app_full_potential}
In an arbitrary coordinate frame, the spatial components of the vector coupling are generally non-zero. This is also the case if one chooses the spatial components to vanish, e.g., in the CMB frame as discussed in \subautoref{sec:frame}. The full resulting potential in the laboratory frame, assuming protons and electrons as the respective particles, is given by (cf. also~\cite{Altschul:2012xu})
\begin{align}
    V_\mathrm{lab}^{pe}(r)
    &=
    \left(g_p^\mathrm{lab}+I_{0,p}^\mathrm{lab}-\boldsymbol I_p^{\:\mathrm{lab}}\cdot\boldsymbol v_p^{\:\mathrm{lab}}\right)\left[g_e^\mathrm{lab}\gamma^0+I_{0,e}^\mathrm{lab}-\boldsymbol I_e^{\:\mathrm{lab}}\cdot\boldsymbol\alpha\right]f(r)\nonumber\\
    &\qquad
    +\frac{1}{2m_p}\varepsilon_{jk\ell}I_{k,p}^\mathrm{lab}\sigma_{\ell,p}\left[g_e^\mathrm{lab}\gamma^0+I_{0,e}^\mathrm{lab}-\boldsymbol I_e^{\:\mathrm{lab}}\cdot\boldsymbol\alpha\right]g_j(\boldsymbol r)\nonumber\\
    &=
    \left(g_p+\gamma_\mathrm{CMB}I_0^p+\gamma_\mathrm{CMB}I_0^p\boldsymbol v_\mathrm{CMB}\cdot\boldsymbol v_p^{\:\mathrm{lab}}\right)\left[g_e\gamma^0+\gamma_\mathrm{CMB}I_0^e+\gamma_\mathrm{CMB}I_0^e\boldsymbol v_\mathrm{CMB}\cdot\boldsymbol\alpha\right]f(r)\nonumber\\
    &\qquad
    -\frac{1}{2m_p}\varepsilon_{jk\ell}\gamma_\mathrm{CMB}I_0^pv_{k,\mathrm{CMB}}\sigma_{\ell,p}\left[g_e\gamma^0+\gamma_\mathrm{CMB}I_0^e+\gamma_\mathrm{CMB}I_0^e\boldsymbol v_\mathrm{CMB}\cdot\boldsymbol\alpha\right]g_j(\boldsymbol r)\label{eq:potential_full}
\end{align}
with $\boldsymbol v_p^{\:\mathrm{lab}}$ the velocity of the proton in the lab frame, $m_p$ the proton mass, $\sigma_{\ell,p}$ the Pauli matrix corresponding to the proton spin, $\boldsymbol\alpha=\gamma^0\boldsymbol\gamma$ and $g_j(\boldsymbol r)$ the derivative of the Yukawa potential function,
\begin{align}
    g_j(\boldsymbol r)
    =
    \partial_jf(r)
    =
    \frac{e^{-m_\phi r}}{4\pi r^2}(1+m_\phi r)\hat r_j,
\end{align}
where $\hat r_j$ is the corresponding component of the radial unit vector. 

It can be shown, term by term, by calculating the corresponding matrix elements that the terms that do not appear in the potential, \autoref{eq:potential_hydrogen},
are subdominant and can hence be dropped to good accuracy. Terms combining $g_F$ or $I_0^F$ with $\boldsymbol{I}_{F'}^\mathrm{lab}$ are parity-odd, such that the energy correction in first-order perturbation theory vanishes. The corresponding leading term in second-order perturbation theory is thus quadratic in the couplings in addition to being velocity-suppressed. The combination of the space-like vector couplings proportional to $f(r)$ is strongly velocity-suppressed. The term in the lower line of \autoref{eq:potential_full}, which combines $\boldsymbol{I}_p^\mathrm{lab}$ and $\boldsymbol{I}_e^\mathrm{lab}$, is a hyperfine-structure effect and hence does not influence our discussion on fine-structure shifts in this work, while also being suppressed by velocity and $m_p$.


\section{Solution of the Dirac Equation for (Anti)hydrogen}
\label{sec:app_Dirac}

We start with the relativistic Dirac equation for the hydrogen atom,
\begin{align}
\label{eq:app_Dirac}
    \left(-i\boldsymbol{\alpha}\cdot\nabla+\beta m_e+V(r)\right)\psi(\boldsymbol{r})=E\psi(\boldsymbol{r})
\end{align}
where $\boldsymbol\alpha$ and $\beta$ are the usual Dirac matrices and $V(r)=-\alpha/r$ is the Coulomb potential. This differential equation can be solved by \cite{rose1961relativistic}
\begin{align}
\label{eq:app_hydr}
    \left(\psi_\kappa^m\right)_\mathrm{H}(\boldsymbol r)
    =
    \begin{pmatrix}
        G(r)\mathcal Y_\kappa^m(\theta,\phi)\\
        iF(r)\mathcal Y_{-\kappa}^m(\theta,\phi)
    \end{pmatrix},
\end{align}
with $G(r)$ and $F(r)$ the real-valued radial functions and $\mathcal Y_\kappa^m(\theta,\phi)$ the angular two-component spinors. The differential equations for the radial and angular parts can be decoupled, leading to the radial equations
\begin{align}
    \left(\frac{\mathrm d}{\mathrm dr}+\frac{1+\kappa}{r}\right)G(r)-\left(m_e+E-V(r)\right)F(r)&=0,\\
    \left(\frac{\mathrm d}{\mathrm dr}+\frac{1-\kappa}{r}\right)F(r)-\left(m_e-E+V(r)\right)G(r)&=0.
\end{align}
For positive-energy particle solutions, $G(r)$ and $F(r)$ are the large and small components, respectively. In order to obtain the solution for antihydrogen, we first take the charge conjugate of \autoref{eq:app_Dirac}, which leads us to the solution of a positron in a repulsive potential.
As the conjugation maps positive-energy electron solutions to negative-energy positron solutions, the physical positron states are not obtained by a simple application of $\psi^c=i\gamma^2\psi^*$. Instead, the radial functions must be redefined accordingly \cite{rose1961relativistic},
\begin{align}
    \left(\psi_\kappa^m\right)_\mathrm{H}^c(\boldsymbol r)
    =
    \begin{pmatrix}
        F^c(r)\sigma_2\left(\mathcal Y_{-\kappa}^m\right)^*\\
        -iG^c(r)\sigma_2\left(\mathcal Y_{\kappa}^m\right)^*
    \end{pmatrix}
\end{align}
Using $\sigma_2\left(\mathcal Y_{\kappa}^m\right)^*=i(-1)^{\ell-j+m}\mathcal Y_{\kappa}^{-m}$, it follows that $G^c(r)$ and $F^c(r)$ are solutions to the same differential equations as $G(r)$ and $F(r)$ with the sign of $V$ inverted \cite{rose1961relativistic}. Moving to antihydrogen, we restore an attractive potential and hence need to invert the sign of the potential again, such that the radial solutions are the same as for hydrogen, but the ordering of the large and small components in the spinor differs,
\begin{align}
    \left(\psi_\kappa^m\right)_{\bar{\mathrm{H}}}(\boldsymbol r)
    =
    \begin{pmatrix}
        -iF(r)\mathcal Y_{-\kappa}^{-m}\\
        G(r)\mathcal Y_{\kappa}^{-m}
    \end{pmatrix}.
\end{align}
Also note that we have dropped an unphysical phase.

This result is consistent with applying the charge conjugation operator to the hydrogen solution from \autoref{eq:app_hydr}, up to a phase. Intuitively, this follows because charge conjugation maps the electron to a positron while simultaneously reversing the sign of the background electromagnetic field, which explains both the form of the angular spinors and the interchange of the large and small components. Thus, the antihydrogen spinor can be understood as the charge-conjugated hydrogen solution.

\section{Statistical Treatment and Correlations}
\label{sec:app_stat}

\subsection{Construction of the Covariance Matrix}
The covariance matrix $\Sigma$ for the differences between experimental and theoretical transition frequencies, given by the entries of the vector $\boldsymbol{\delta}$, contains three contributions: (i) experimental, (ii) pure theoretical, and (iii) input-parameter-induced uncertainties. While the experimental and pure theoretical covariance matrices contain correlations between some transitions, the contribution arising from the uncertainties of the input constants in the theoretical calculation leads to correlations among all transitions. We provide a short overview of these contributions and give the resulting correlations at the end.

Unless measurements originate from the same experiment, we assume the corresponding transition frequencies to be uncorrelated. This leads to a small number of non-vanishing correlations in the experimental contribution (i). Since the $2\mathrm{S}_{1/2}-4\mathrm S_{1/2}$ transition frequency is obtained by combining the quantity $(E_{4\mathrm S_{1/2}}-E_{2\mathrm S_{1/2}})-\frac 1 4(E_{2\mathrm S_{1/2}}-E_{1\mathrm S_{1/2}})$ with the measurement of the $1\mathrm{S}_{1/2}-2\mathrm S_{1/2}$ transition frequency, correlation with the $1\mathrm{S}_{1/2}-2\mathrm S_{1/2}$ measurement is introduced through shared experimental input. For the $2\mathrm{S}_{1/2}-4\mathrm P_{j}$ transitions, we use the correlations given in the supplemental material of \cite{Beyer:2017gug}. Similarly, we use the Methods section of \cite{Maisenbacher:2026nau} to extract a covariance matrix for the $2\mathrm{S}_{1/2}-6\mathrm P_{j}$ transitions. In both cases, the hyperfine corrections are consistently included in the covariance matrix. For the $1\mathrm{S}_{1/2}-2\mathrm S_{1/2}$ transitions of antihydrogen, we use Table 3 from \cite{Baker:2025ehs} to infer the correlations among the three measurements. Correlations also arise from the magnetic-field and hyperfine-splitting corrections following \cite{Rasmussen:2017pyn}. Lastly, we consider the Lamb shift and $2\mathrm P$ fine-structure splittings for antihydrogen from \cite{ALPHA:2020rbx}. Assuming vanishing correlation between the primary inputs $\Delta E(2\mathrm S,2\mathrm{P_c})$ and $\Delta E(2\mathrm S,2\mathrm{P_f})$, a correlation coefficient of $r\approx 0.939$ is obtained after uncertainty propagation using the formulae in the Methods section. 

We use the uncertainties and correlation coefficients from Tables XII and XIII of \cite{Mohr:2024kco} to determine the pure theoretical covariance matrix (ii). As described in \subautoref{sec:theo_values}, the theoretical uncertainties for the $6\mathrm P_j$ states are not provided. By comparing with the other $n\mathrm P_j$ states, we assume $\sigma_\mathrm{th}(2\mathrm S_{1/2})\gg\sigma_\mathrm{th}(6\mathrm P_j)$, such that the uncertainty of the $6\mathrm P_j$ states is subdominant in the $2\mathrm S_{1/2}-6\mathrm P_j$ transitions. Since the $4\mathrm P_j$ states show no correlations with other states, we assume the same to hold approximately for the $6\mathrm P_j$ states.

The contribution to the covariance matrix from input-parameter-induced uncertainties (iii) is obtained by propagating the uncertainties of the input constants \cite{CODATA2022data} through the theoretical expressions for the transition frequencies \cite{Mohr:2024kco}. This contribution induces correlations among all transitions through their common dependence on the input constants.

The diagonal elements of the covariance matrix follow from the experimental and combined purely theoretical and input-constant-induced uncertainties listed in \autoref{tab:data} by quadratic addition, while the off-diagonal elements are constructed from the correlation matrix given in \autoref{tab:correlation_matrix}, using the index mapping given in the first column of \autoref{tab:data}. We point out that the correlations of transitions 14 and 15 with all other transitions except for their mutual correlation are numerically negligible due to their comparatively large experimental uncertainties.

\begin{table}[t]
\setlength{\tabcolsep}{3pt}
    \centering
    \resizebox{\columnwidth}{!}{%
    \begin{tabular}{c | c c c c c c c c c c c c c c c}
    \hline\hline
         & 1&2&3&4&5&6&7&8&9&10&11&12&13&14&15\\
        \hline
         1&1 & 0.990 & -0.028 & -0.007 & 0.100 & 0.321 & 0.321 & 0.853 & 0.862 & 0.528 & 0.596 & 0.570 & 0.426 &0&0\\
         2& &1&-0.026 & -0.007 & 0.010 & 0.319 & 0.319 & 0.849 & 0.858 & 0.525 & 0.590 & 0.564 & 0.422&0&0\\
         3& & &1&0.001 & -0.001 & -0.006 & -0.006 & -0.013 & -0.013 & -0.007 & -0.017 & -0.016 & -0.012&0&0\\
         4& & & &1&0 & -0.001 & -0.001 & -0.003 & -0.003 & -0.001 & -0.004 & -0.004 & -0.003&0&0\\
         5& & & & &1&0.033 & 0.033 & 0.089 & 0.090 & 0.055 & 0.060 & 0.057 & 0.043&0&0\\
         6& & & & & &1&0.116 & 0.283 & 0.286 & 0.175 & 0.192 & 0.183 & 0.137&0&0\\
         7& & & & & & &1&0.283 & 0.286 & 0.175 & 0.192 & 0.183 & 0.137&0&0\\
         8& & & & & & & &1&0.753 & 0.469 & 0.509 & 0.486 & 0.364&0&0\\
         9& & & & & & & & &1&0.474 & 0.514 & 0.491 & 0.368&0&0\\
         10& & & & & & & & & &1&0.315 & 0.301 & 0.225&0&0\\
         11& & & & & & & & & &  &1&0.532 & 0.398&0&0\\
         12& & & & & & & & & & & &1&0.393&0&0\\
         13&&& & & & & & & & & & &1&0&0\\
         14&&& & & & & & & & & & & &1&0.939\\
         15&&& & & & & & & & & & & & &1\\
         \hline\hline
    \end{tabular}
    }
    \caption{Correlation matrix for the hydrogen and antihydrogen transitions considered in the analysis. The transitions corresponding to the indices are listed in \autoref{tab:data}. Entries with $\vert r(i,j)\vert<0.001$ are set to zero for clarity.}
    \label{tab:correlation_matrix}
\end{table}

\subsection{Least-Squares Fit}
Since the energy corrections discussed in \subautoref{sec:energy_correction} depend linearly on the four products $g_p^\pm g_e^\pm$, the least-squares problem has an exact solution that can be written in a closed form \cite{Cowan:1998ji}. Introducing the parameter vector $\boldsymbol\theta$, whose components correspond to the four products, the energy correction for a given transition can be written as
\begin{align}
    \Delta_i(\boldsymbol\theta)
    =
    \sum_{j=1}^4A_{ij}\theta_j,
\end{align}
where $i$ corresponds to a specific transition between two quantum states and the matrix $A$ encodes the structure of the energy corrections for the considered quantum states, in particular the radial integrals. The $\chi^2$ function is thus given by
\begin{align}
    \chi^2
    =
    (\boldsymbol{\delta}-\boldsymbol{\Delta}(\boldsymbol{\theta}))^T\Sigma^{-1}(\boldsymbol{\delta}-\boldsymbol{\Delta}(\boldsymbol{\theta}))
    =
    (\boldsymbol{\delta}-A\boldsymbol{\theta})^T\Sigma^{-1}(\boldsymbol{\delta}-A\boldsymbol{\theta}),
\end{align}
where $\Sigma$ denotes the covariance matrix of $\boldsymbol\delta$. Minimizing the $\chi^2$ function yields the best-fit point of the parameters,
\begin{align}
    \boldsymbol{\tilde{\theta}}
    =
    \tilde\Sigma A^T\Sigma^{-1}\boldsymbol{\delta},
\end{align}
where we have defined $\tilde\Sigma\equiv (A^T\Sigma^{-1}A)^{-1}$. Expanding around the best-fit point, $\boldsymbol\theta=\boldsymbol{\tilde\theta}+\boldsymbol\xi$, the $\chi^2$ function can be written as
\begin{align}
    \chi^2
    =
    \chi_\mathrm{min}^2+\boldsymbol{\xi}^T\tilde\Sigma^{-1}\boldsymbol{\xi}
\end{align}
with the covariance matrix of the fitted parameters given by $\tilde\Sigma$. Confidence regions are obtained from contours of constant $\Delta\chi^2=\chi^2-\chi_\mathrm{min}^2$.

\bibliographystyle{utphys}
\bibliography{references}

\end{document}